\documentstyle{amsppt}
\newcount\mgnf\newcount\tipi\newcount\tipoformule\newcount\greco 
\tipi=2          
\tipoformule=0   

\global\newcount\numsec\global\newcount\numfor
\global\newcount\numapp\global\newcount\numcap
\global\newcount\numfig\global\newcount\numpag
\global\newcount\numnf

\def\SIA #1,#2,#3 {\senondefinito{#1#2}%
\expandafter\xdef\csname #1#2\endcsname{#3}\else
\write16{???? ma #1,#2 e' gia' stato definito !!!!} \fi}

\def \FU(#1)#2{\SIA fu,#1,#2 }

\def\etichetta(#1){(\veroparagrafo.\veraformula)%
\SIA e,#1,(\veroparagrafo.\veraformula) %
\global\advance\numfor by 1%
\write15{\string\FU (#1){\equ(#1)}}%
\write16{ EQ #1 ==> \equ(#1)  }}
\def\etichettaa(#1){(A\veraappendice.\veraformula)
 \SIA e,#1,(A\veraappendice.\veraformula)
 \global\advance\numfor by 1
 \write15{\string\FU (#1){\equ(#1)}}
 \write16{ EQ #1 ==> \equ(#1) }}
\def\getichetta(#1){Fig. \verafigura
 \SIA g,#1,{\verafigura}
 \global\advance\numfig by 1
 \write15{\string\FU (#1){\graf(#1)}}
 \write16{ Fig. #1 ==> \graf(#1) }}
\def\retichetta(#1){\numpag=\pgn\SIA r,#1,{\verapagina}
 \write15{\string\FU (#1){\rif(#1)}}
 \write16{\rif(#1) ha simbolo  #1  }}
\def\etichettan(#1){(n\verocapitolo.\veranformula)
 \SIA e,#1,(n\verocapitolo.\veranformula)
 \global\advance\numnf by 1
\write16{\equ(#1) <= #1  }}

\newdimen\gwidth
\gdef\profonditastruttura{\dp\strutbox}
\def\senondefinito#1{\expandafter\ifx\csname#1\endcsname\relax}
\def\BOZZA{
\def\alato(##1){
 {\vtop to \profonditastruttura{\baselineskip
 \profonditastruttura\vss
 \rlap{\kern-\hsize\kern-1.2truecm{$\scriptstyle##1$}}}}}
\def\galato(##1){ \gwidth=\hsize \divide\gwidth by 2
 {\vtop to \profonditastruttura{\baselineskip
 \profonditastruttura\vss
 \rlap{\kern-\gwidth\kern-1.2truecm{$\scriptstyle##1$}}}}}
\def\verapagina{
{\romannumeral\number\numcap}.\number\numsec.\number\numpag}}

\def\alato(#1){}
\def\galato(#1){}
\def\veroparagrafo{\number\numsec}\def\veraformula{\number\numfor}
\def\veraappendice{\number\numapp}
\def\verapagina{\number\pageno}\def\veranformula{\number\numnf}
\def\verafigura{{\romannumeral\number\numcap}.\number\numfig}
\def\verocapitolo{\number\numcap}\def\veranformula{\number\numnf}
\def\Eqn(#1){\eqno{\etichettan(#1)\alato(#1)}}
\def\eqn(#1){\etichettan(#1)\alato(#1)}
\def\ver{\veroparagrafo}
\def\Eq(#1){\eqno{\etichetta(#1)\alato(#1)}}
\def\eq(#1){\etichetta(#1)\alato(#1)}
\def\Eqa(#1){\eqno{\etichettaa(#1)\alato(#1)}}
\def\eqa(#1){\etichettaa(#1)\alato(#1)}
\def\dgraf(#1){\getichetta(#1)\galato(#1)}
\def\drif(#1){\retichetta(#1)}

\def\eqv(#1){\senondefinito{fu#1}$\clubsuit$#1\else\csname fu#1\endcsname\fi}
\def\equ(#1){\senondefinito{e#1}\eqv(#1)\else\csname e#1\endcsname\fi}
\def\graf(#1){\senondefinito{g#1}\eqv(#1)\else\csname g#1\endcsname\fi}
\def\rif(#1){\senondefinito{r#1}\eqv(#1)\else\csname r#1\endcsname\fi}
\def\bib[#1]{[#1]\numpag=\pgn
\write13{\string[#1],\verapagina}}

\def\include#1{
\openin13=#1.aux \ifeof13 \relax \else
\input #1.aux \closein13 \fi}

\openin14=\jobname.aux \ifeof14 \relax \else
\input \jobname.aux \closein14 \fi
\openout15=\jobname.aux
\openout13=\jobname.bib


\ifnum\tipoformule=1\let\Eq=\eqno\def\eq{}\let\Eqa=\eqno\def\eqa{}
\def\equ{}\fi


{\count255=\time\divide\count255 by 60 \xdef\hourmin{\number\count255}
        \multiply\count255 by-60\advance\count255 by\time
   \xdef\hourmin{\hourmin:\ifnum\count255<10 0\fi\the\count255}}

\def\oramin{\hourmin }

\def\data{\number\day/\ifcase\month\or january \or february \or march \or
april \or may \or june \or july \or august \or september
\or october \or november \or december \fi/\number\year;\ \oramin}

\setbox200\hbox{$\scriptscriptstyle \data $}

\newcount\pgn \pgn=1
\def\foglio{\number\numsec:\number\pgn
\global\advance\pgn by 1}
\def\foglioa{A\number\numsec:\number\pgn
\global\advance\pgn by 1}



\global\newcount\numpunt

\magnification=\magstephalf
\baselineskip=16pt
\parskip=8pt

\voffset=2.5truepc
\hoffset=0.5truepc
\hsize=6.1truein
\vsize=8.4truein 
{\headline={\ifodd\pageno\rightheadline \else \leftheadline \fi}}
\def\rightheadline{\it  {Section $\ver$}\hfil\tenrm\folio}
\def\leftheadline{\tenrm \folio \hfil\it  {Gaussian Hopfield}}

\def\a{\alpha}
\def\b{\beta}
\def\d{\delta}

\def\f{\phi}

\def\s{\sigma}

\def\th{\theta}

\def\z{\zeta}
\def\o{\omega}

\def\O{\Omega}

\def\1{{1\kern-.25em\hbox{\rm I}}}
\def\eu{{1\kern-.25em\hbox{\sm I}}}
\def\f1{{1\kern-.25em\hbox{\vsm I}}}

\def\R{{\Bbb R}}  
\def\N{{\Bbb N}}  
\def\P{{\Bbb P}}  
\def\Q{{\Bbb Q}}  
\def\E{{\Bbb E}}  



\let\cal=\Cal
\def\AA{{\cal A}}
\def\BB{{\cal B}}

\def\DD{{\cal D}}

\def\FF{{\cal F}}

\def\SS{{\cal S}}

\def\NN{{\cal N}}
\def\MM{{\cal M}}

\def\QQ{{\cal Q}}
\def\PP{{\cal P}}

\def\chap #1#2{\line{\ch #1\hfill}\numsec=#2\numfor=1}

\def\sb{{\subset}}
\def\wt{\widetilde}

\def\limlaw{\buildrel \DD\over\rightarrow}

\def\note#1{\footnote{#1}}

\def\frac#1#2{{#1\over #2}}

\def\text#1{\quad{\hbox{#1}}\quad}
\def\newpage{\vfill\eject}
\def\proposition #1{\noindent{\thbf Proposition #1:}}

\def\theo #1{\noindent{\thbf Theorem #1: }}
\def\lemma #1{\noindent{\thbf Lemma #1: }}

\def\corollary #1{\noindent{\thbf Corollary #1: }}
\def\proof{{\noindent\pr Proof: }}
\def\proofof #1{{\noindent\pr Proof of #1: }}
\def\endproof{$\diamondsuit$}
\def\remark{\noindent{\bf Remark: }}
\def\thanks{\noindent{\bf Acknowledgements: }}
\font\pr=cmbxsl10
\font\thbf=cmbxsl10 scaled\magstephalf

\font\ch=cmbx12
\font\ftn=cmr8

\font\it=cmti10
\font\bf=cmbx10
\font\sm=cmr7
\font\vsm=cmr6


\overfullrule=0pt

\font\tit=cmbx12
\font\aut=cmbx12
\font\aff=cmsl12
\def\s{\char'31}
\centerline{\tit STOCHASTIC SYMMETRY-BREAKING IN A}
\vskip.2truecm
\centerline{\tit GAUSSIAN HOPFIELD-MODEL}
\vskip.2truecm 
\vskip1.5truecm

\centerline{\aut Anton Bovier 
\note{ e-mail:
bovier\@wias-berlin.de} 
}
\vskip.1truecm
\centerline{\aff Weierstra\s {}--Institut}
\centerline{\aff f\"ur Angewandte Analysis und Stochastik}
\centerline{\aff Mohrenstrasse 39, D-10117 Berlin, Germany}
\vskip.4truecm
\centerline{\aut  Aernout C.D. van Enter\note{\ftn
e-mail: A.C.D.van.Enter\@phys.rug.nl}}
\vskip.1truecm
\centerline{\aff Institute for Theoretical Physics}
\centerline{\aff Rijksuniversiteit Groningen}
\centerline{\aff Nijenborgh 4}
\centerline{\aff NL-9747 AG Groningen, The Netherlands}
\vskip.4truecm
\centerline{\aut  Beat Niederhauser\note{
e-mail: niederha\@math.tu-berlin.de}${}^,$\note{
supported by DFG in the Graduiertenkolleg ``Stochastische Prozesse und 
probabilistische Analysis''}}
\vskip.1truecm
\centerline{\aff Fachbereich Mathematik}
\centerline{\aff Technische Universit\"at Berlin}
\centerline{\aff Strasse des 17. Juni 136}
\centerline{\aff D-10623 Berlin, Germany}

\vskip1truecm\rm
\def\s{\sigma}
\noindent {\bf Abstract:} We study a ``two-pattern'' Hopfield model
with Gaussian disorder. We find that there are infinitely many pure states
at low temperatures in this model, and we find that the metastate is 
supported on an infinity of symmetric pairs of pure states. The origin of this 
phenomenon is the random breaking of a rotation symmetry of the distribution of
the disorder variables.


$ {} $
\newpage

\chap{1. Introduction: Ising spins with a rotation symmetry}1

In this paper we will illustrate the notions of chaotic size dependence, metastates
and their dispersal, and the chaotic pairs of states scenario, introduced
as a possible description of the low temperature spin glass phase [N,NS2,NS3,NS4,NS6],
on a simple model which is similar to the two-state Hopfield model.
The fact that the model has site disorder makes it
more tractable than the commonly considered bond-disorder spin glass models.
The main difference with the standard Hopfield model of neural networks,
is that instead of two i.i.d. Bernoulli random variables
the disorder is described by  two i.i.d. Gaussian random 
variables at every site. 
As a consequence, in the thermodynamic limit we obtain the existence, for a 
``two-pattern'' model, of uncountably many (instead of two times two) pure states for 
this model, due to the existence
of a continuous (rotation) symmetry of the distribution of the random
variables describing the disorder. In any finite volume, however, this 
symmetry is necessarily
randomly broken in a given realization. Intuitively, this means that there are only two pure 
ground states, and  the low temperature Gibbs state is close to the symmetric 
mixture of two, out
of a possible continuum, of pure Gibbs states, due to the
fluctuations in the disorder. 

The concepts we want to illustrate
have their origin in the theory of spin-glasses. However, the most often 
considered spin-glass models, which have
bond disorder, both in finite dimension (the Edwards-Anderson models)
and the equivalent neighbour (Sherrington-Kirkpatrick) model, have turned
out to be so complicated to analyze, that up till now it has not been possible
to check which of the possible scenarios  for the spin-glass phase applies to 
them. 

We remind the reader that in the debate within 
the physics literature on the extreme sides there are the proposals
of Fisher and Huse, [FH1,FH2,FH3,FH4] predicting the existence of 
only two pure states in any 
dimension higher or equal than 3, versus the 
proposal of Parisi and coworkers, in which an infinity of pure states is
predicted [MPV, MPR]. This scenario has been claimed to apply down to 
the 3-dimensional Edwards-Anderson model.
Intermediate scenarios have been discussed 
by [BF,NS1,NS2,NS3,NS4,NS5,NS6,N,vE]. 

Although of course lattice models 
with two pure states are common, our experience with models having an infinite 
number of pure states is a lot more limited. Therefore we hope that our 
discussion will be useful in illustrating various concepts, mostly 
introduced and studied in a systematic way by Newman and Stein 
(see in particular [N,NS2,NS3,NS4,NS6]), which have been 
introduced either in an abstract setting or via (in)formal arguments, 
by applying them to  a concrete model.

The main idea in the approach of Newman and Stein 
is to classify the possible scenarios on the basis of first principles,
using only general ergodic properties
using the concept of ``metastates'', i.e. 
probability distributions on the space of Gibbs 
measures (first introduced apparently in [AW]; see [N,NS2,NS3,Ku1,Ku2,BG3] for
more details, as well as applications of these concepts 
and extensions to equivalent 
neighbour or mean-field type models--to which our model also belongs).  


In this context, in one of their most 
recent papers [NS6], they conjectured that in a disordered 
lattice system, in any approximate decomposition of a finite volume Gibbs 
states into ``pure states'', the weights in this decomposition should
be mostly concentrated on a single subset of states that are 
related by an exact symmetry of the system, while  other states 
would appear with a weight that tends to 
zero as the volume tends to infinity. The particular subset chosen 
could of course be random and could depend strongly on the volume. 
Applied to the Ising spin glass situation,
this argument would predict the chaotic pairs picture.


Although a similar 
situation has been shown to occur in the usual Hopfield model 
with $M=\a N$ patterns if $\a$ is small in [BG3],  we found it worthwhile to 
construct a simple model showing these features in order to see what is 
involved. 

Let us state the definitions of our variant of the Hopfield model and the main
quantities of interest. Let $\SS_N = \{-1,+1\}^N$ denote the set of functions
$\s : \{1,\ldots,N\} \rightarrow \{-1,+1\}$, and the set $\SS = \{-1,+1\}^{\N}$.
We call $\s$ a spin configuration and denote by $\s_i$ the value of $\s$ at
$i$. Let $(\O, \FF, \P)$ be an abstract probability space and let 
$\xi^{\mu}_i[\o]$, $i \in \N$, $\mu=1,2$, denote a family of i.i.d.\ 
standard Gaussian variables. We will write $\xi^{\mu}[\o]$ for the 
$N$-dimensional vector whose $i$th component is given by $\xi^{\mu}_i[\o]$;
such a vector is called a {\it pattern}. On the other hand, we will write 
$\xi_i[\o]$ for the two dimensional vector with the same components. When
we write $\xi[\o]$ without indices, we consider it as a $2 \times N$
matrix (its transpose will be denoted by $\xi^t$).

Throughout the paper, $(\cdot,\cdot)$ denotes the scalar product, without indication
of the space where its arguments lie.

We define random maps $m^{\mu}_N[\o](\s): \SS_N \rightarrow [-1,+1]$
(conventionally called {\it overlap parameters})
through
$$
m^{\mu}_N[\o](\s) \equiv \frac{1}{N} \sum_{i=1}^N \xi^{\mu}_i[\o]\s_i.
\Eq(I.1)
$$
The Hamiltonian
is now defined as
$$
\eqalign{
H_N[\o](\s) &\equiv - \frac{N}{2} \sum_{\mu=1,2} \Big(m^{\mu}_N[\o](\s)\Big)^2 \cr
&= - \frac{N}{2} \|m_N[\o](\s)\|_2^2,
}
\Eq(I.2)
$$
where $\| \cdot \|_2$ denotes the $l_2$-norm in $\R^2$.

Note that if we rewrite $\xi'{}^1_{i} =
\xi^{\theta}{}^1_{i} = \xi^1_{i} \cos(\theta) + 
\xi^2_{i} \sin(\theta)$
and $\xi'{}^2_{i} = \xi'{}^{\theta}_{i} = \xi^1_{i} \sin(\theta) - \xi^2_{i} 
\cos(\theta)$ 
the Hamiltonian 
has the same form in the primed variables. However, this transformation 
is a {\it statistical} symmetry, mapping one disorder realization of the model
to another one, drawn from the same distribution, as opposed to for example the
spin-flip symmetry which is an exact symmetry for any given realization of 
the disorder.

Through this Hamiltonian, finite volume Gibbs measures on $\SS_N$ are defined by
$$
\mu_{N,\b}[\o](\s) \equiv 2^{-N} \frac{e^{-\b H_N[\o](\s)}}{Z_{N,\b}[\o]},
\Eq(I.3)
$$
and the induced distribution of the overlap parameters
$$
\QQ_{N,\b}[\o] \equiv \mu_{N,\b}[\o] \circ m_N[\o]^{-1}. 
\Eq(I.4)
$$

The normalizing factor in \eqv(I.3), called the {\it partition function}, 
is explicitly given by
$$
Z_{N,\b}[\o] \equiv 2^{-N} \sum_{\s \in \SS_N}e^{-\b H_N[\o](\s)} 
\equiv \E_\s\,e^{-\b H_N[\o](\s)}.
\Eq(I.5)
$$
We are mainly interested in the concentration behaviour of $\QQ_{N,\b}$ as 
$N \rightarrow \infty$. 
It will be convenient to do this by considering the auxiliary measure 
$\wt{\QQ}_{N,\b}\equiv{\QQ}_{N,\b}\star\NN_2(0, \frac{1}{\b N}\1)$
obtained by a convolution with a Gaussian measure, 
its so-called Hubbard-Stratonovich transform. 
Since, for $N$ large, $\NN_2(0, \frac{1}{\b N}\1)$ converges rapidly to 
the Dirac measure at zero, the two measures have asymptotically the
same properties. For details see e.g. [BGP]. 
$\wt{\QQ}_{N,\b}$ is  absolutely continuous with respect to Lebesgue measure
on $\R^2$  and has
the density 
$$
\frac{ e^{-\b N\Phi_{N,\b}[\o](z)} }{Z_{N,\b}[\o]}, 
\Eq(I.6)
$$
where $\Phi_{N,\b}$ is given by
$$
\Phi_{N,\b}[\o](\s) = \frac{1}{2} \|z\|_2^2 - \frac{1}{\b N} \sum_{i=1}^N
\ln\,\cosh \b(\xi_i[\o],z).
\Eq(I.7)
$$

As usual in  mean-field models, we construct the 
extremal Gibbs measures by {\it tilting} the Hamiltonian \eqv(I.2) with an
{\it external magnetic field} (for a general discussion on the 
issue of limiting Gibbs states in mean field models, see [BG1], Sect. 2.4 or
[BG3], Sect. 2). That is, we define
a more general Hamiltonian
$$
H_N^h[\o](\s) \equiv - \frac{N}{2} \|m_N[\o](\s)\|_2^2 - N(h,m),
\Eq(I.7bis)
$$
where $h= (b \cos(\vartheta), b \sin(\vartheta)) \in \R^2$. The corresponding
measures on the spins and on $\R^2$ are denoted by $\mu_{N,\b}^h[\o]$ and
$\QQ_{N,\b}^h[\o]$, respectively. We then take the limits 
$\lim_{b \rightarrow 0} \lim_{N \rightarrow \infty}$, for all values
of $\vartheta \in [0,2\pi)$. 
We distinguish the measures constructed from this Hamiltonian by an additional 
superscript $h$.

We are now able to give a precise formulation of our main results.

\theo {1}
{\it 
  Let ${\bf h} = (b \cos\vartheta, b \sin \vartheta)$. Then
$$
\lim_{b \rightarrow 0} \lim_{N \rightarrow \infty} \QQ_{N,\b}^{\bf h}
  = \delta_{(r^* \cos \vartheta, r^* \sin \vartheta)},
\Eq(I.8)
$$
where $r^*$ is the largest solution of the equation
$$
r^* = \frac 1{\sqrt{2\pi}}\int dx\, e^{-\frac{x^2}{2}}x\tanh(\b x r^* ).
\Eq(I.9)
$$
}

Theorem 1 shows that there is an uncountable number of extremal limiting 
induced measures, indexed by the circle. The following Corollary shows 
that to each of them corresponds a distinct 
 limiting Gibbs measure
on the spins.

\corollary {2}
{\it
  For any finite set $I \subset \N$, and $\P$-almost all $\omega$,
$$
\mu_{\infty,\b}^h[\o]\Big(\{\s_I = s_I\}\Big) \equiv
        \lim_{b \rightarrow 0} \lim_{N \rightarrow \infty} \mu_{N,\b}^h[\o] 
        \Big(\{\s_I = s_I\}\Big) = \prod_{i \in I} 
        {e^{\b s_i (\xi_i[\o], m)} \over
        2 \cosh(\b (\xi_i[\o], m))},
\Eq(I.10)
$$	
where $m = (r^* \cos(\vartheta), r^* \sin(\vartheta))$, and $r^*$
as in \eqv(I.9).
}

In Theorem 1 and Corollary 2 convergence is 
almost sure due to the presence of the 
tilting field.  The situation changes if we set $b=0$ first and take 
the infinite volume limit later.

\theo {3}
{\it
  Let $\QQ_{N,\b}$ as in \eqv(I.4) and $m=m(\vartheta)=(r^* \cos\vartheta, r^* \sin \vartheta)$,
where $\vartheta \in [0,\pi)$ is a uniformly distributed random variable. Then
$$
\QQ_{N,\b} \limlaw \frac{1}{2} \delta_{m(\vartheta)}
        + \frac{1}{2} \delta_{-m(\vartheta)} \equiv \QQ_{\infty,\b}[m].
\Eq(I.11)
$$
Furthermore, the (induced)  
AW-metastate is the image of the uniform distribution of $\vartheta$ 
under the measure-valued map $\vartheta \mapsto \QQ_{\infty,\b}[m(\vartheta)]$.
}

\corollary {4}
{\it
Let $I \subset \N$ be finite. Then the following holds:
\item{(i)} Let $\{g_i\}_{i \in I}$ be a family of i.i.d.\ random variables, distributed
as $\NN(0,r^*)$. Then
$$\displaystyle \lim_{N \uparrow \infty}\mu_{N,\b}(\s_I = s_I)
        \limlaw
\frac{1}{2} \prod_{i \in I}\frac{e^{\b s_i g_i}}{2 \cosh \b g_i} +
\frac{1}{2} \prod_{i \in I}\frac{e^{-\b s_i g_i}}{2 \cosh \b g_i} .
\Eq(I.12)
$$
\item{(ii)} The AW-metastate is the image of the 
uniform distribution on $\vartheta$ under the measure-valued map  
$\vartheta \mapsto \mu_{\infty,m(\vartheta)}[\o]$ where
$$
\mu_{\infty,\b,m}[\o] =
\frac{1}{2} \prod_{i \in I}\frac{e^{\b s_i (\xi_i[\o],m)}}{2 \cosh \b (\xi_i[\o],m)} +
\frac{1}{2} \prod_{i \in I}\frac{e^{-\b s_i (\xi_i[\o],m)}}{2 \cosh \b (\xi_i[\o],m)}.
\Eq(I.13)
$$
}
Statement (ii) of Corollary 4 motivates the notion of metastates. Whereas on the level of
the induced measures $\QQ_{N,\b}$ one cannot see any influence by the conditioning, this
is clearly the case on the level of the Gibbs measures on the spins. 

The remainder of this paper is mainly devoted to the proofs of the 
two theorems (the corollaries are standard consequences (see e.g. 
[BGP1] or [BG3] for proofs of analogous statements in more complicated 
situation) and will not be given) 
is organized as follows. In Section 2 we prove 
the necessary concentration estimates on the measures $\QQ_{N,\b}$. 
This will yield immediately Theorem 1. In the case $h=0$ we will show that
the measure concentrates near the absolute minima of some
random process, and in Section 3 we will analyse the properties of these
minima. In particular we will prove that these converge 
in distribution to one-point sets.  This will allow us to prove Theorem 3. 
In Section 4 we discuss some further consequences on the chaotic volume 
dependence, the empirical metastate and the superstate.

\remark We consider the case of two  patterns here in order to keep 
technicalities to a minimum. All our results can be extended without 
any novel difficulties to the case of any fixed finite number, $M$, 
of Gaussian 
patterns. In that case the set of extremal Gibbs measures will be indexed by 
the sphere in $\R^M$ and the metastate will be supported on pairs 
of mirror images on this sphere, with the position being uniformly 
distributed. Thus nothing really new will happen. The situation 
when the number of patterns grows with the volume may be more interesting 
and work in this direction is in progress.

\newpage
\chap{2. Concentration}2
\def\oo{a}

In this section we show the concentration properties of the measures 
$\wt\QQ_N$ for large $\b$. These imply the same concentration 
results for the measures $\QQ_N$ by standard arguments that have been
developed in much more complicated situations, see e.g. [BG2]. 
The estimates presented here are mostly similar, and often much simpler,
to those that can be found e.g. in [BG2], but we decided to present some
parts in detail where some care is required.  


We start with the more delicate case $h=0$ that will be relevant for
the proof of Theorem 3 (which will be given at the end of Section 
3). 
We are interested in the concentration behaviour of the measures 
$\wt{\QQ}_{N,\b}$. The following two lemmata each give a partial
answer. The first one asserts that $\wt{\QQ}_{N,\b}$ is concentrated
exponentially about a circle around the origin, whereas the second one
tells us that even on this circle, only a small part really contributes
to the total mass.

\lemma {\ver.1}
 {\it 
  Let $\{\xi_i^{\mu}\}_{i \in {\N}, \mu = 1,2}$ be i.i.d.\ standard
  Gaussian variables, and define $\Phi_{N,\b} (z)$ as
  $$
  \Phi_N (z) \equiv \frac{1}{2} \|z\|_2^2 - \frac{1}{\b N}
      \sum_{i=1}^N \ln \, \cosh \beta (\xi_i ,z).
  \Eq(C.80)       
  $$

  Let furthermore $\delta_N = N^{-1/10}$.
  Then there exist strictly positive constants $K,K'$, $m,m'$ 
  such that ($r^*$ is the largest solution in \eqv(I.9))
  $$              
  \frac{\int_{|\,\|z\| - r^*| \geq \delta_N} e^{-\b N \Phi_N(z)}
    \, dz}{\int_{|\,\|z\| - r^*| < \delta_N} e^{-\b N \Phi_N(z)}
    \, dz} \leq K e^{-K N^{m}},
  \Eq(C.7)
$$
on a set of ${\P}$-measure at least $1-K'e^{-K'N^{m'}}$.
}

The second result needs an additional definition. Let 
$$
g_N(\vartheta)   \equiv \frac{1}{\sqrt{N}}
        \sum_{i=1}^N \ln\,\cosh(\b r^*\zeta_i 
        \cos(\vartheta-\varphi_i)),
\Eq(C.101)
$$
where $(\zeta_i, \varphi_i)$ are the polar coordinates of the two
dimensional vector $\xi_i$.

\lemma {\ver.2}
{\it 
  Assume the hypotheses of Lemma \ver.1. Let $\oo_N = N^{- 1/25}$.
  Then there exist strictly positive constants $K_1,K_2,C_1,C_2$
  such that on a set of ${\P}$-measure at least
  $$
    1-K_1 e^{-N^{-1/25}}
  \Eq(C.40)
  $$
  the following bound holds,
  $$
    \frac{\int_{A'_N} e^{-\b N \Phi_N(z)}\, dz}{\int_{A_N}
                 e^{-\b N\Phi_N(z)}\, dz}
    \leq C_1  e^{-N^{2/5}},
  \Eq(C.41)
  $$
  where 
  $$
    \eqalign{
      A_N &= \Big\{(r,\vartheta) \in {\R}^+_0 \times 
        [0,2\pi) \big| |r-r^*| < \delta_N, g_N(\vartheta)
          - \min_{\vartheta} g_N(\vartheta) < \oo_N \Big\}, \cr
      {A'}_N &= \Big\{(r,\vartheta) \in {\R}^+_0 \times 
         [0,2\pi) \big| |r-r^*| < \delta_N,  g_N(\vartheta)
          - \min_{\vartheta} g_N(\vartheta) \geq \oo_N \Big\}.
    }
  \Eq(C.42)
  $$
}
Combining these two lemmata and using the Borel-Cantelli lemma, 
we get immediately  the following result.

\proposition {\ver.3}
{\it  
  Assume the hypotheses of Lemma 2.1.
  Then there exist strictly positive constants $K,K',m$, such that
  $$
 \P\left[ 
\frac{\int_{A_N^c} e^{-\b N\Phi_N(z)}\, dz}{\int_{A_N}e^{-\b N\Phi_N(z)}\,
      dz} > K e^{-K'N^m},{\tenrm \ i.o.\ in\ } N\right]=0,
  \Eq(C.81)
  $$
  where $A_N$ is as in Lemma \ver.2.
}

To see why the preceding results should be expected, we must consider the 
function $\Phi_{N,\b} $.
Note that the  expectation of this function,
$$
    {\E}\, \Phi_N(z) = \frac{1}{2} \|z\|_2^2 - 
\frac{1}{\b}{\E}\,\ln\,\cosh\beta(\xi_1,z).
  \Eq(C.4)
$$
depends only on the modulus of its argument. It is useful to observe that
if $z=(r \cos\th, r\sin\th)$, we can represent $\E\,\Phi_N(z)$ as
$$
    \E\,\Phi_N(z)= \frac 12 r^2-\E_\varphi\E_{\,\zeta} 
 \ln\,\cosh
      (\beta r \zeta \cos(\varphi)) \, d\varphi
    \Eq(C.6)
$$
where $\zeta,\phi$ are the representation of the polar decomposition 
of a two dimensional normal vector, i.e. $\zeta$ is distributed
with density $xe^{-x^2/2}$ on ${\R}^+$, and $\varphi$ uniformly on 
the circle $[0,2\pi)$.

From this it follows that $\E\,\Phi_N(z)$ takes its minimum 
on the circle with radius $r^*(\b)$, where $r^*$ is defined in Theorem 1.
It is easy to verify that there is $0<\b^*<\infty$, such that 
$r^*(\b)>0$ if and only if $\b>\b^*$.  

It is also straightforward to check that ${\E}\, \Phi$ is 
sufficiently smooth to guarantee that
it is bounded from above by a quadratic function (of $\|z\|$) in some 
neighbourhood containing $r^*$.

\proofof {\ver.1} We start with the numerator. We decompose the domain of 
integration into an ``inner'' part ${\cal I}$,and an ``outer'' part ${\cal O}$:
$$
\eqalign{
  \left\{z \in {\R}^2: \left| \|z\| - r^* \right| \geq 
       \delta \right\}  
     &= \left\{z \in {\R}^2: \|z\| - r^* \geq
          \delta \right\} \cr
     & \quad \cup  \left\{z \in {\R}^2: 
          \|z\| - r^* \leq - \delta \right\} 
     = {\cal O} \cup {\cal I}.
     }
   \Eq(C.8)
$$
Consider the integral on ${\cal O}$. We write it as
$$
   \int_{\cal O} e^{-N \Phi_N(z)}\, dz = \int_{\cal O} 
      e^{- \b N {\E}\, \Phi_N(z)} e^{-\b N(\Phi_N(z) -
        {\E}\,\Phi_N(z))} \, dz,
      \Eq(C.9)
$$
and observe that ${\E}\,\Phi_N$ can be bounded below by a 
quadratic function $C(\|z\| - r^*)^2$. We are left with the task
of estimating the term $\Phi_N(z) - {\E}\,\Phi_N(z)$. This
is accomplished by the following Lemma.

\lemma {\ver.4}
{\it Let $f_N(z) = \frac{1}{\b N}\sum_{i=1}^N \ln\,\cosh \beta(\xi_i,z)$
and 
$$
 {\cal O} = \left\{z \in {\R}^2:\|z\| > r^* + \delta
\right\}.
\Eq(C.10)
$$
Then, for $\delta$ small enough, such that $\d^2/16 \leq \d/2\sqrt 2$, 
there exist strictly positive constants  $C_1,C_2,K_1,K_2$ such that
$$
  {\P}\left[\sup_{z \in {\cal O}} \left| f_N(z) -
     {\E}f_N(z) \right| \geq \frac{C}{2}(\|z\| - r^*)^2 \right]
  \leq K_1 e^{-K_2N} + C_1\delta^{-2}e^{-C_2 \delta^4 N}N^{-\frac{1}{2}}.
\Eq(C.11)
$$
}
\proof Define $\bar{f}_N(z) = f_N(z) - {\E}\,f_N(z)$. The left-hand side
of \eqv(C.11) is bounded from above by
$$
\eqalign{
  &\leq {\P}
  \left[ \sup_{z' \in {\cal W}_r \cap {\cal O}}
    \left| \bar{f}_N(z') \right| \geq
    \frac{C}{4}\left(\|z'\| - r^* \right)^2 
  \right] \cr
  & \quad + {\P}\left[ \sup_{z' \in {\cal W}_r \cap A}
    \sup_{z \in B_r(z')} \left| \bar{f}_N(z) - \bar{f}_N(z') \right|
    \geq \frac{C}{4} \left( \|z'\| - r^* \right)^2 \right],
  }
\Eq(C.12)
$$
where ${\cal W}_r$ is the grid with spacing $r$ in ${\R}^2$, 
and $z' \in {\cal W}_r$ is chosen such that 
$0 \leq \|z\| - \|z'\| < \sqrt{2}\,r$.

The argument of the second term can be uniformly bounded. 
Using e.g. Lemma 6.10 of [BG1], we get that 
$$
    | f_N(z) - f_N(z')| 
    \leq \|z-z'\|_2 \|A\|^{1/2},
\Eq(C.13)
$$
where $A$ is the matrix $(1/N)\xi^T \xi$. Similarly, 
$$
  | {\E}\, f_N(z) - {\E}\, f_N(z')| \leq \|z-z'\|_2 ({\E} \|A\|)^{1/2}.
\Eq(C.14)
$$
Now, a trivial computation shows that
$$
\E\,\|A\|\leq 1+C/\sqrt N
\Eq(C.014)
$$
and using (for instance) the same argument as in Section 4 of [BG1],
but replacing Talagrand's concentration estimate for bounded r.v.'s
by the standard Gaussian concentration inequality (see e.g. [LT], Ch. 1),
one shows easily that 
$$
\P\left[|\|A\|-1|\geq x\right]\leq Ce^{-Nx^2/C}.
\Eq(C.015)
$$
Therefore, 
$$
  \eqalign{
    {\P} \Bigg[  \sup_{z' \in {\cal W}_r \cap A} 
      \sup_{z \in B_r(z')} |\bar{f}_N(z) &- \bar{f}_N(z') | \geq 
      \frac{C}{4} (\|z'\| - r^*)^2] \Bigg] \cr
    &\leq {\P} \left[ r(\|A\|^{1/2} + ({\E}
      \|A\|)^{1/2}) \geq \frac{C}{4}(\|z'\| - r^*)^2 \right] \cr
    &\leq {\P} \left[ (\|A\|^{1/2} + 2)
      \geq \frac{C\d^2}{4r} \right],
  }
\Eq(C.15)
$$
Choosing the grid parameter $r$ such that $r \leq C\delta^2/16$
 the right-hand side of \eqv(C.15)
is bounded by
$
  {\P} \left[ \|A\|>4\right]\leq Ce^{-9N/C}
$
This takes care of the second term in \eqv(C.12).
Let us now treat the first term. The probability that the 
supremum over all lattice points of some function exceeds some given value 
is transformed into  a summable series of probabilities that at each lattice point 
the function is greater than this value. More precisely, we have
$$
  \eqalign{
    {\P}
     \left[ \sup_{z' \in {\cal W}_r \cap {\cal O}}
       \left|\bar{f}_N(z') \right| \geq
       \frac{C}{4}\left(\|z'\| - r^* \right)^2  \right] 
     &\leq \sum_{z' \in {\cal W}_r \cap {\cal O}}
     {\P} \left[ \left|\bar{f}_N(z') \right| \geq 
       \frac{C}{4}\left(\|z'\| - r^* \right)^2  \right] \cr
     &\leq \sum_{z' \in {\cal W}_r \cap {\cal O}}
       e^{-K C^2\left(\|z'\| - r^*\right)^4 N},
     }
\Eq(C.18)
$$
by Chebyshev's inequality. Then
$$
  \eqalign{ \sum_{z' \in {\cal W}_r \cap {\cal O}}
    e^{-K C^2\left(\|z'\| - r^*\right)^4 N}
     &= r^{-2} \sum_{z' \in {\cal W}_r \cap {\cal O}}
        r^2 e^{-K C^2\left(\|z'\| - r^*\right)^4 N}\cr
     &\leq r^{-2} \int_{{\R}^2 \backslash B_0(r^* +
      \delta - \sqrt{2}\,r)} e^{-K C^2
        \left(\|z'\| - r^*\right)^4 N} \, dz \cr
     &\leq r^{-2} e^{-K \frac{C^2}{16}\delta^4 N}
        \int_{{\R}^2 \backslash B_0(r^* + \delta / 2)}
        e^{-K \frac{C^2}{16} \left(\|z'\| - r^*\right)^4 N}
        \, dz \cr
    &\leq r^{-2} 2\pi e^{-K \frac{C^2}{16}\delta^4 N} N^{-\frac{1}{2}}
       \int_{\delta/2}^{\infty}z e^{-\wt{K} z^4} \, dz \cr
    &\leq K' r^{-2} e^{-K \frac{C^2}{2}\delta^4 N} N^{-\frac{1}{2}},
  }
\Eq(C.19)
$$
where $K'$ stands for an upper bound for the integral, which is 
independent of $N$ (assuming $\delta > 2\sqrt{2}r$).
 Combining this and \eqv(C.15),
and choosing $\delta$ small enough such that $C\d^2/16 \leq \d/(2 \sqrt{2})$
concludes the proof of Lemma \ver.4.
\endproof

Therefore, on a set of measure at least $1 - C_1 e^{-C_2 N\delta^4}$,
the integral \eqv(C.9) can be bounded by 
$$
  \eqalign{
    \int_{\cal O}e^{-\b N{\E}\,\Phi_N(z)} e^{-\b N
      \left(\Phi_N(z) - {\E}\,\Phi_N(z) \right)} \, dz
    & \leq \int_{\cal O}e^{-\b N\frac{C}{2} \left(\|z\| - r^*
      \right)^2} \, dz \cr
    & \leq 2 \pi \int_{r^* + \delta}^{\infty} r 
       e^{- \b N C (r - r^*)^2} \, dz \cr
    & \leq 2 \pi e^{- N \frac{C}{4} \delta^2} \int_0^{\infty}
       r e^{- \b N \frac{C}{4} r^2} \, dr \cr
    &= 2 \pi \frac{2}{\b N C} e^{- \b N \frac{C}{4} \delta^2} .
  }
\Eq(C.20)
$$
We now turn to the integral on the ``inner'' part $\cal I$.
Again, we have to control the term
$$
  \Phi_N(z) - {\E}\,\Phi_N(z).
\Eq(C.21)
$$
Since $\cal I$ is compact, we can do this uniformly by using 
the following lemma.

\lemma {\ver.5}
{\it  Let $f_N(z) = 1/(\b N)\sum_{i=1}^N \ln \, \cosh \beta(\xi_i,z)$
and $A \subset {\R}^2$ a bounded set. Then there exist
strictly positive constants $K_1,K_2,C_1,C_2$ such that
$$
  {\P}\left[\sup_{z \in A}|f_N(z) - {\E}\, f_N(z)|
    > \varepsilon \right] \leq K_1 e^{-K_2 N} + C_1 \varepsilon^{-2}
  e^{-C_2 \varepsilon^2 N}.
\Eq(C.22)
$$
}
The proof is similar (if not simpler) to the proof of Lemma \ver.4 and
is left to the reader.\endproof

Lemma \ver.5 implies that
$$
  \eqalign{
    \int_{\cal I}e^{- \b N \Phi_N(z)} \, dz 
    &\leq e^{\varepsilon N} e^{-\b N {\E} \, \Phi(r^*)}
      \int_{\cal I}e^{- \b N {\E}\, \Phi_N(z)} \, dz \cr
    &\leq e^{\varepsilon \ b N}e^{-\delta^2 C \b N} \pi r^*{}^2,
  }
\Eq(C.29)
$$
using the fact that ${\E}\,\Phi_N(\|z\|) -  
{\E} \, \Phi(r^*)$ can be bounded
uniformly on $\cal I$ by its value for $\|z\| = r^* - \delta$.

Finally, the denominator in \eqv(C.7) can be bounded
from below, using the second order Taylor expansion with remainder of 
${\E}\,\Phi_N(\|z\|)$
$$
  \eqalign{
&    \int_{\left|\, \|z\| - r^* \right| < \delta} \!\!
      e^{- \b N \Phi_N(z)}\, dz\cr 
    &\geq e^{-\b N {\E} \, \Phi(r^*)}\!
    \int_{\left|\, \|z\| - r^* \right| < \delta} \!\!
        e^{- N C \left( \|z\| - r^*\right)^2 - N C' \left(
       \|\tilde{z}\| - r^* \right)^3 -N \varepsilon } \, dz \cr
    &\geq 2 \pi \frac{1}{\b NC}e^{-\varepsilon \b N} e^{-\b NC'\delta^3}
       e^{-\b N{\E} \, \Phi(r^*)}
       \left(1 - \delta e^{-\b N C \delta^2}\right),
  }
\Eq(C.30)
$$
on a set of measure at least 
$1 - K e^{-K N}- C \varepsilon^{-2}e^{-C N\varepsilon^2}$ 
(this error term can be estimated by Lemma \ver.5).
Collecting \eqv(C.20), \eqv(C.29) 
and \eqv(C.30), we get that on a set of measure 
exponentially close to one,
$$
  \eqalign{
    \frac{\int_{\left| \|z\| - r^* \right| \geq \delta}
             e^{- \b N \Phi_N(z)}\, dz}
      {\int_{\left| \|z\| - r^* \right| < \delta}
        e^{- \b N \Phi_N(z)}\,dz} 
    &\leq M e^{\varepsilon \b N}e^{\b NC'\delta^3}(2 \pi)^{-1} \b NC 
      \left(1 - \delta e^{-\b NC \delta^2} \right)^{-1} \cr
    &\quad \times \left\{e^{\varepsilon \b N}e^{-\b N C \delta^2} 
        \pi r^*{}^2 + 2 \pi e^{-\b N \frac{C}{4}\delta^2}
            \frac{2}{\b NC} \right\} \cr
    &= MK e^{-\b N(C \delta^2 - 2\varepsilon -C'\delta^3)}\b N \left(
      1 - \delta e^{-NC\delta^2}\right)^{-1} \cr
    &\quad + M K' e^{-\b N(\frac{C}{4} \delta^2 - \varepsilon - 
        C' \delta^3)} N \left(1 - \delta e^{-\b NC\delta^2}
        \right)^{-1}.
  }
\Eq(C.31)
$$
%
Now let us choose
$\delta_N = N^{-\frac{1}{10}}$, $\varepsilon_N = N^{-\frac{1}{4}}$; 
then \eqv(C.31) gives
$$
  \eqalign{
    \frac{\int_{|\,\|z\| - r^*| \geq \delta_N} e^{-\b N \Phi_N(z)}
    \, dz}{\int_{|\,\|z\| - r^*| < \delta_N} e^{- \b N \Phi_N(z)}
    \, dz}
    &\leq M \wt{K}Ne^{-\b N^{\frac{4}{5}}(C - 2 
      N^{-\frac{1}{20}} - C' N^{-\frac{1}{10}})} \cr
    &\quad + M \wt{K}Ne^{- N^{\frac{4}{5}}(\frac{C}{4}
      - N^{-\frac{1}{20}} - C' N^{-\frac{1}{10}})},
  }
\Eq(C.38)
$$
on a set which is exponentially close (in $N$) to 1. This 
concludes the proof of Lemma \ver.1.
\endproof

We now turn to the proof of Lemma \ver.2 which is a little more delicate than 
the previous one. 

\proofof {\ver.2}
Let us write $I(B)$ for the integral $\int_B e^{-\b N\Phi_N(z)}
\, dz$.
We will prove the concentration behaviour by a strategy similar to the
one used in Lemma \ver.1. Namely we replace
the function $\Phi_N$ by its expectation ${\E}\,\Phi_N$
and control the error. 

Write the fluctuation term $\Phi_N - {\E}\,\Phi_N$ as
$$
  \eqalign{
    \Phi_N(z) - {\E}\,\Phi_N(z) &=
      \frac{1}{\b N} \sum_{i=1}^N \{\ln\,\cosh \beta (\xi_i,z)
          - {\E}\,\ln\,\cosh \beta (\xi_i,z)\} \cr
    &= \frac{1}{\b N} \sum_{i=1}^N \{\ln\,\cosh \beta (\xi_i,z)
      - \ln\,\cosh \beta (\xi_i,z') \cr
    & \quad\quad\quad - {\E}\,\ln\,\cosh \beta (\xi_i,z)
       + {\E}\,\ln\,\cosh \beta (\xi_i,z')\} \cr
    & \quad + \frac{1}{\b N} \sum_{i=1}^N \{\ln\,\cosh \beta 
         (\xi_i,z') - {\E}\,\ln\,\cosh \beta (\xi_i,z')\}.
  }
\Eq(C.43)
$$
Now choose  $z'$ such that $z' = z'(z)=\lambda z$, $\lambda > 0$, 
and $\|z'\| = r^*$ (i.e.\ $z'$ is the projection of $z$ onto 
$S^1(r^*)$). Define the two functions
$$
  \eqalign{
    h_N(z) &\equiv 
    \frac{1}{\sqrt{N}} \sum_{i=1}^N \{\ln\,\cosh \beta (\xi_i,z)
      - \ln\,\cosh \beta (\xi_i,z') \cr
    & \quad\quad\quad - {\E}\,\ln\,\cosh \beta (\xi_i,z)
       + {\E}\,\ln\,\cosh \beta (\xi_i,z')\},
  }
\Eq(C.44)
$$
with $z'$ defined as above, and
$$
  g_N(z) \equiv \frac{1}{\sqrt{N}} \sum_{i=1}^N 
           \{\ln\,\cosh \beta (\xi_i,z)
          - {\E}\,\ln\,\cosh \beta (\xi_i,z)\}.
\Eq(C.45)
$$
Then the fluctuation term takes the form
$$
  N(\Phi_N(z) - {\E}\,\Phi_N(z)) = 
      \frac{\sqrt{N}}{\b} (h_N(z) -  g_N(z')).
\Eq(C.46)
$$

It is the term $g_N$ that determines the concentration behaviour of the
measure. To see this we first 
bound the term $h_N$ uniformly on the ``annulus of concentration'' 
$A_N \cup A'_N$. We have
the following result.

\lemma {\ver.6}
  {\it Let $\{\xi_i\}_{i \in {\N}}$ be i.i.d.\  Gaussian variables
with mean zero and variance one. Let $h_N$ be as in \eqv(C.44), and 
$A_N, A'_N$ as in \eqv(C.42). Then for any $\varepsilon > 0$,
$$
  {\P}\left[ \sup_{z \in A_N \cup A'_N} |h_N(z)| \geq
    \varepsilon \right] \leq K N^2 e^{- N^{1/10}(\varepsilon - K N^{-1/10})}.
\Eq(C.47)
$$
}
\proof Let us write 
$$
  f_i(z) \equiv \ln\,\cosh\beta(\xi_i,z), 
\Eq(C.48)
$$
and
$$
  \bar{f}_i \equiv \ln\,\cosh\beta(\xi_i,z) - {\E}\, 
     \ln\,\cosh\beta(\xi_i,z).
\Eq(C.49)
$$
We also keep the notation $z' = z'(z)$ defined above. Introduce a polar grid
${\cal W}_N$ in ${\R}^2$, i.e.\ a discrete set of points $x_{i,j}$ whose 
polar coordinates are given by $(\rho_i,\alpha_j) \in {\R}^+ \times [0,2\pi)$,
such that $\Delta_N \alpha \equiv |\alpha_i - \alpha_j| = K N^{-1/2}$ and
$\Delta_N \rho \equiv |\rho_i - \rho_j| = K N^{-1/2}$, for some appropriate
constant $K$. Note that for any point $z$ in a bounded domain 
$A \subset {\R}^2$, the distance to the closest grid point
is less than $K' N^{-1/2}$. 

For any $z \in {\R}^2$, define $x = x(z) \in {\cal W}_N$ to be the
grid point closest to $z$, and $y = y(z) \in {\cal W}_N$ the grid point closest
to $z' = z'(z)$. One can easily convince oneself, that $x' = y'$, i.e.\ the
two points $x$ and $y$ lie on the same ray starting at the origin. Then we 
can decompose the function $h_N(z)$ as
$$
  \eqalign{
    h_N(z) &= \frac{1}{\sqrt{N}} \sum_{i=1}^N \{\bar{f}_i(z) - \bar{f}_i(z') \} \cr
    &= \frac{1}{\sqrt{N}} \sum_{i=1}^N \{\bar{f}_i(z) - \bar{f}_i(x)\} 
    + \frac{1}{\sqrt{N}} \sum_{i=1}^N \{\bar{f}_i(x) - \bar{f}_i(y)\} \cr
    &\quad + \frac{1}{\sqrt{N}} \sum_{i=1}^N \{\bar{f}_i(y) - \bar{f}_i(z')\}.
  }
\Eq(C.50)
$$
Denote by $I_1(z,x)$, $I_2(x,y)$, $I_3(y,z')$ respectively the first,
second and third sum on the right-hand side of \eqv(C.50).
We can then write (let ${\cal A}_N  = A_N \cup A'_N$, the ``annulus of 
concentration'')
$$
  \eqalign{
    {\P} \left[ \sup_{z \in {\cal A}_N} |h_N(z)| 
      \geq \varepsilon \right] 
    &=  {\P} \left[ \sup_{z \in {\cal A}_N} |I_1(z,x) 
      + I_2(x,y) + I_3(y,z')| \geq \varepsilon \right] \cr
    &\leq {\P} \left[ \sup_{x \in {\cal W}_N \cap {\cal A}_N}
      \sup_{z \in B_{K N^{-1/2}}(x)} |I_1(z,x)| \geq 
      \frac{\varepsilon}{3}\right]  \cr
    &\quad +  {\P} \left[ \sup_{x \in {\cal W}_N \cap {\cal A}_N} 
      \sup_{y \in  {\cal W}_N \cap {\cal A}_N \atop
        y' = x'}
      |I_2(x,y)| \geq \frac{\varepsilon}{3}\right] \cr
    &\quad + {\P} \left[ \sup_{y \in {\cal W}_N \cap {\cal A}_N} 
      \sup_{z' \in B_{K N^{-1/2}}(y)} |I_3(y,z')| \geq
      \frac{\varepsilon}{3}\right].
  }
\Eq(C.51)
$$
The first and the third term (they are equal) can be uniformly bounded 
by an estimate analogous to the proof of Lemma \ver.2. In fact, for 
any $u,v$, we have
$$
  \eqalign{
  \left| \frac{1}{\sqrt{N}} \sum_{i=1}^N\{\bar{f}_i(u) - \bar{f}_i(v)\} \right|
  &\leq \sqrt{N} \beta ( \|A\|^{1/2} + ({\E}\, \|A\|)^{1/2})
    \|u - v\|_2.
  }
\Eq(C.52)
$$
Now, if $\|u - v\|_2 \leq 4 \varepsilon' N^{-1/2} / \beta$, we have the 
following exponential bound.
$$
  \eqalign{
    {\P} \left[ | \frac{1}{\sqrt{N}} \sum_{i=1}^N
      \{\bar{f}_i(u) - \bar{f}_i(v)\} | \geq \varepsilon'\right] 
    &\leq {\P} \left[ \|A\|^{1/2} + ({\E}\, \|A\|)^{1/2}
      \geq \frac{\varepsilon' N^{- 1/2}}{\beta \|u-v\|_2} \right] \cr
    &\leq {\P} \left[ \|A\| \geq 4 \right] \leq K e^{-K N}.
  }
\Eq(C.53)
$$
Thus we get for the first term 
in \eqv(C.51),
$$
  \eqalign{
   & {\P}\Bigg[ \sup_{x \in {\cal W}_N \cap {\cal A}_N}
      \sup_{z \in B_{K N^{-1/2}}(x)} |I_1(z,x)| \geq 
      \frac{\varepsilon}{3}\Bigg]  \cr
    &\leq \sum_{x \in {\cal W}_N \cap {\cal A}_N} {\P}
      \left[ \sup_{z \in B_{K N^{-1/2}}(x)} |I_1(z,x)| \geq 
      \frac{\varepsilon}{3}\right] \cr
&    \leq \sum_{x \in {\cal W}_N \cap {\cal A}_N} {\P}
      \left[ \|A\| \geq 4 \right] 
    \leq K N^{1/10} N^{-1} e^{- K N},
  }
\Eq(C.54)
$$
since we know that $\|x - z\| = K' N^{-1/2}$, by the remark preceding
\eqv(C.50), and the number of grid points in ${\cal A}_N$ is bounded
by $N^1 \delta_N^{-1}$ times some constant. The same estimate is valid
for the term containing $I_3$ (since they are equal). 

Let us now consider the term containing $I_2$. We know that 
$\|x - y\| \leq 2 \delta_N$, since those two points are supposed
to lie on the same ``ray''. Again, we can turn the supremum into
a sum,
$$
  {\P} \left[ \sup_{x \in {\cal W}_N \cap {\cal A}_N}
    \sup_{{y \in  {\cal W}_N \cap {\cal A}_N} \atop
      {y' = x'}} |I_3(x,y)| \geq \frac{\varepsilon}{3}\right]
  \leq \sum_{x,y} {\P} \left[ |I_3(y,z')| \geq 
    \frac{\varepsilon}{3}\right],
\Eq(C.55)
$$
where $x,y$ on the right-hand side satisfy the same conditions as on
the left-hand side. By Chebyshev's inequality, we get that for any
$u$, $v$
$$
  \eqalign{
    {\P}\left[\sum_{i=1}^N \{\bar{f}_i(u) -\bar{f}_i(v)\}
      \geq \sqrt{N} \varepsilon'\right]
    &\leq \inf_{s >0} e^{-s\varepsilon'\sqrt{N}}
      {\E}\left[e^{s\sum_{i=1}^N \{\bar{f}_i(u) -\bar{f}_i(v)\}}
        \right] \cr
    &= \inf_{s >0} e^{-s\varepsilon'\sqrt{N}}\prod_{i=1}^N
      {\E} e^{s \{ \bar{f}_i(u) -\bar{f}_i(v)\}} .
    }
\Eq(C.56)
$$
Now we use the series expansion of the exponential function, the fact
that the exponent in the right-hand side of \eqv(C.56) is a centered
random variable, and some obvious inequalities for each term of 
the expansion, to get
$$
  {\E} e^{s \{ \bar{f}_i(u) -\bar{f}_i(v)\}} 
  \leq \left\{1 + \frac{s^2}{2}{\E}\left[(\bar{f}_i(u) -\bar{f}_i(v))^2
      e^{s|\bar{f}_i(u) -\bar{f}_i(v)|}\right] \right\}.
\Eq(C.57)
$$
To evaluate the expectation term, we use the inequality
$$
  |f_i(u) - f_i(v)| \leq \beta |(\xi_i,u-v)|.
\Eq(C.58)
$$
Then the expectation term in \eqv(C.56) is bounded by
$$
  \eqalign{
   {\E}\left[(\bar{f}_i(u) -\bar{f}_i(v))^2 
     e^{s|\bar{f}_i(u) -\bar{f}_i(v)|}\right]
   &\leq \left({\E}\left[(\bar{f}_i(u) -\bar{f}_i(v))^4\right]
     \right)^{\frac{1}{2}}\left({\E}\,e^{2s|\bar{f}_i(u) -\bar{f}_i(v)|}
       \right)^{\frac{1}{2}} \cr
   &\leq 4 \left({\E}\left[(f_i(u) - f_i(v))^4\right]\right)^{\frac{1}{2}}
     \left({\E}\,e^{2s|f_i(u) - f_i(v)|}\right)^{\frac{1}{2}} \cr
   &\quad \times e^{s{\E}\,|f_i(u) - f_i(v)|}\;,
  }
\Eq(C.60)
$$
where the first inequality follows by Cauchy-Schwarz, and the second one
is a consequence of the inequality $(a+b)^2 \leq 2(a^2 + b^2)$ (applied twice
to the first factor), respectively the trivial fact that $|a-b| \leq |a| + |b|$.
All quantities in \eqv(C.60) can be bounded easily using \eqv(C.58). One
gets (by calculating explicit Gaussian integrals)
$$
  \eqalign{
    {\E} \left[(f_i(u) - f_i(v))^4 \right] 
    &= 3\|u-v\|_2^4,
  }
\Eq(C.61)
$$
$$
  \eqalign{
    {\E}\,e^{2s|f_i(u) - f_i(v)|} 
    &\leq 2e^{2s^2\|u-v\|_2^2},
  }
\Eq(C.62)
$$
$$
  \eqalign{
    e^{s {\E}\,|f_i(u) - f_i(v)|}
    &\leq e^{s\sqrt{2/\pi} \|u-v\|_2}.
  }
\Eq(C.63)
$$
Inserting \eqv(C.61)--\eqv(C.63) into \eqv(C.60), gives
$$
  \frac{s^2}{2}{\E}\left[(\bar{f}_i(u) -\bar{f}_i(v))^2 
     e^{s|\bar{f}_i(u) -\bar{f}_i(v)|}\right] 
   \leq 2 \sqrt{6} s^2 \|u-v\|_2^2 e^{2 s^2\|u-v\|_2^2 + s\sqrt{2/\pi}
     \|u-v\|_2}.
\Eq(C.64)
$$
We use the above bound \eqv(C.64) in \eqv(C.56), together with the 
inequality $1 + x \leq e^x$, and the fact that $\|x-y\|_2 \leq \delta_N =
K N^{-1/10}$. We thus get the following estimate
$$
  {\P} \left[\sum_{i=1}^N \{\bar{f}_i(u) - \bar{f}_i(v)\}
    \geq \sqrt{N} \varepsilon' \right] 
  \leq \inf_{s > 0} e^{-s\varepsilon'\sqrt{N} + K s^2 N^{4/5}
    e^{2 s^2 N^{-1/5} + \sqrt{2/\pi} N^{-1/10}}}.
\Eq(C.65)
$$
Choosing $s = N^{-2/5}$, this gives 
$$
  {\P} \left[\sum_{i=1}^N \{\bar{f}_i(u) - \bar{f}_i(v)\}
    \geq \sqrt{N} \varepsilon' \right] 
  \leq \wt{K} e^{-N^{1/10}(\varepsilon' - K N^{-1/10})}.
\Eq(C.66)
$$
The same bound applies to 
$$
  {\P}\left[\sum_{i=1}^N \{\bar{f}_i(u) - \bar{f}_i(v)\}
    \leq - \sqrt{N} \varepsilon' \right].
\Eq(C.67)
$$
Inserting \eqv(C.66) and \eqv(C.67) into the left-hand side 
of \eqv(C.55) gives
$$
  {\P}\left[\sup_{x \in {\cal W}_N \cap {\cal A}_N}
    \sup_{y \in  {\cal W}_N \cap {\cal A}_N \atop
      y' = x'} |I_2(x,y)| \geq \varepsilon' \right]
  \leq K N^{1/2}N^{1/10} e^{-N^{1/10}(\varepsilon' - K' N^{-1/10})},
\Eq(C.68)
$$
since the number of terms in the sum does not exceed a constant times
$N^{1/2}$ (the number of allowed $x$) times $N^{1/10}$ (the number of
allowed $y$).
Using \eqv(C.54) and \eqv(C.68),  \eqv(C.51) gives
$$
  {\P}\left[\sup_{z \in {\cal A}_N} | h_N(z)| \geq \varepsilon
    \right] 
  \leq K N^2 e^{-K'N^{1/10} \varepsilon}.
\Eq(C.69)
$$
This concludes the proof of Lemma \ver.6.
\endproof

Note that we can choose $\varepsilon$ as a function of $N$, and still
get an exponential bound. For example, choose 
$\varepsilon = \varepsilon_N \equiv (\ln N)^2 N^{-1/20}$. Lemma
\ver.6 then reads

\lemma {\ver.7}
{\it Let $\{\xi_i\}_{i \in {\N}}$ be i.i.d.\ Gaussian variables
with mean zero and variance one. Let $h_N$ be as in \eqv(C.44),
and $A_N,A'_N$ as in \eqv(C.42). Then,
$$
  {\P} \left[ \sup_{z \in A_N \cup A'_N} |h_N(z)| 
    \geq N^{-1/20} (\ln N)^2 \right] 
  \leq K N^2 e^{-N^{1/20}((\ln N)^2 - K' N^{-1/20})}.
\Eq(C.70)
$$
Furthermore, 
$$
  {\P} \left[ \sup_{z \in A_N \cup A'_N} |h_N(z)|
    \geq N^{-1/20} (\ln N)^2, {\tenrm i.o.\ in\ } N 
        \right] = 0.
\Eq(C.71)
$$
}
\proof The first statement (equation \eqv(C.70)) is a 
straightforward consequence of Lemma \ver.6. Equation \eqv(C.71)
then follows by the first Borel-Cantelli Lemma.
\endproof


Let us now estimate the integral $I({A'}_N)$. We get explicitly,
using the  bound on $h_N$ from Lemma 
\ver.6,
$$
  \eqalign{
    \int_{{A'}_N} e^{-\b N \Phi_N(z)} \, dz 
    &= \int_{{A'}_N} e^{-\b N {\E}\,\Phi_N(z)}
       e^{-\sqrt{N} h_N(z)}e^{-\sqrt{N} g_N(z'(z))}\, dz \cr
    &\leq \int_{|r-r^*|<\delta_N}  r e^{-\b N {\E}\,
       \Phi_N(r^*)} e^{\sqrt{N}\varepsilon} \, dr \cr
    &\quad \times \int_{g_N(\vartheta)-\min g_N
        > \oo_N} e^{-\sqrt{N} g_N(\vartheta)} \, 
      d\vartheta \cr
    &= 2 e^{- \b N {\E}\, \Phi_N(r^*)}
       e^{\sqrt{N} \varepsilon}\int_{|r-r^*|<\delta_N} r\,dr \cr
    &\quad \times \int_{g_N(\vartheta) - \min g_N> \oo_N} 
        e^{-\sqrt{N} g_N(\vartheta)} \, d\vartheta \cr
    &\leq 4 e^{- \b N {\E}\, \Phi_N(r^*)}
         e^{\sqrt{N} \varepsilon}r^*\delta_N\cr
       &\quad \times 2 \pi e^{-\sqrt{N} \oo_N} e^{- \sqrt{N} \min g_N}.
  }
\Eq(C.73)
$$
Thus,
$$
  \int_{{A'}_N} e^{- \b N \Phi_N(z)} \, dz  \leq
    K e^{- \b N {\E}\, \Phi_N(r^*)}
    e^{\sqrt{N} \varepsilon} 
    \delta_Nr^* e^{-\sqrt{N} \oo_N}.
\Eq(C.76)
$$
We now turn to the integral $I(A_N)$. Using standard
estimates for Gaussian integrals, a quadratic upper bound
of $g_N$ about its minima, and the fact that
${\E}\, \Phi(\|z\|)$ can be bounded from above by  a quadratic
function in some neighbourhood containing $r^*$, we get
$$
  \eqalign{
    \int_{A_N} e^{- \b N\Phi_N(z)}\, dz &\geq
      e^{- \b N {\E}\, \Phi_N(r^*)}  e^{- \sqrt{N}\varepsilon}
      \int_{|r-r^*|<\delta_N} r e^{- \b NC'(r-r^*)^2} \, dr \cr
    &\quad \times \int_{g_N(\vartheta) - \min g_N
      \leq \oo_N} e^{-\sqrt{N} g_N(\vartheta)} \,
      d\vartheta \cr
    &\geq K e^{- \b N {\E}\, \Phi_N(r^*)} 
      e^{-\sqrt{N} \varepsilon}(r^* - \delta_N)
      \left(\frac{\pi}{NC'}\right)^{1/2} \cr
    &\quad (1- e^{-NC'\delta_N})\left(\frac{\pi}{K\sqrt{N}}
      \right)^{1/2}(1-e^{-\sqrt{N}K' \oo_N}).
  }
\Eq(C.77)
$$
We get finally for the ratio $I({A'}_N)/I(A_N)$
$$
    \frac{I(A'_N)} {I(A_N)} 
    \leq K \frac{r^*}
    {r^* - \delta_N} N^{3/4} e^{-\sqrt{N}(\oo_N - 2 \varepsilon)}.
\Eq(C.78)
$$
Lemma \ver.7 allows us to choose $\varepsilon = \varepsilon(N) = N^{-1/20} (\ln N)^2$.
Inserting this choice, together with $\oo_N = N^{-1/25}$, into
\eqv(C.78), gives
$$
  \frac{I(A'_N)} {I(A_N)} \leq K N^{3/4} e^{-N^{23/50}(1 - K'(\ln N)^2
    N^{-1/100})}.
\Eq(C.79)
$$
This statement is true for all $\omega \in \Omega$, for which Lemma \ver.6
respectively \ver.7 holds, that is on a set of ${\P}$-measure
at least $K N^2 e^{-N^{1/20}((\ln N)^2 - K' N^{-1/20})}$. This proves  
Lemma \ver.2. 
\endproof

Let us now turn to the proof of Theorem 1. We again state first a result 
about the concentration of the induced measure $\wt{\QQ}_{N,\b}^h$.

\proposition {\ver.8}
{\it
  Let $\{ \xi_i^{\mu}\}_{i \in \N, \mu = 1,2}$ be i.i.d.\ standard Gaussian
variables, and define
$$
 \Phi_{N,\b}^h (z) \equiv \frac{1}{2} \|z\|_2^2     
        - \frac{1}{\b N} \sum_{i=1}^N \ln \, \cosh \b (\xi_i , z + h).
\Eq(C.201)
$$
Let furthermore $\d_N = N^{-1/5}$. Then there exist strictly positive constants
$K,K',m$ such that
$$
\P \Bigg\{ \frac{\int_{\|z -\tilde{r}^h\| \geq \d_N} e^{-\b N \Phi_{N,\b}^h(z)}\, dz}
        {\int_{\|z -\tilde{r}^h\| < \d_N} e^{-\b N \Phi_{N,\b}^h(z)}\, dz}
\geq Ke^{-K' N^m}, \tenrm{\ i.o.\ in\ } N \Bigg\} = 0,
\Eq(C.202)
$$
where $\tilde{r}^h$ is the unique minimum of the function
$$
\E\, \Phi_{N,\b}^h(z) = \frac{1}{2} \|z\|_2^2 - \frac{1}{\b} \E \ln \, \cosh
        \b(\xi_1, z+h).
\Eq(C.203)
$$
}

\proof 
Let us decompose $\Phi_{N,\b}^h$ in the usual way
$$
\Phi_{N,\b}^h(z) = \E \Phi_{N,\b}^h(z) + \Phi_{N,\b}^h(z) - \E \Phi_{N,\b}^h(z).
\Eq(C.204)
$$
We first treat the denominator appearing in \eqv(C.202). $\E \Phi_{N,\b}^h$ can be
bounded from below by some quadratic function $C\|z-\tilde{r}^h\|_2^2$ on the 
set $\|z -\tilde{r}^h\| \geq \d_N > 0$. The fluctuation term can be controlled
by the following analogue of Lemma \ver.4.

\lemma {\ver.9}
{\it 
  Let $f_N = \frac{1}{\b N} \sum_{i=1}^N \ln\,\cosh \b (\xi_i, z+h)$. Then for 
$\d$ small enough, sucht that $C\d^2/80 < \d/2$, there exist strictly positive
constants $C_1,C_2,K_1,K_2$ such that
$$
\eqalign{
p_N \equiv \P \Big[ \sup_{z:\|z - \tilde{r}^h\|_2 \geq \d}
        |f_N(z) - \E f_N(z)|  & \geq \frac{C}{2} \|z - \tilde{r}^h\|_2^2 \Big] \cr
&\leq K_1 e^{-K_2 N} + C_1 N^{1/2}\d^{-2} e^{-C_2 N}.
}
\Eq(C.205)
$$
}

\proof 
The proof is completely analogous to the proof of Lemma \ver.4, and is left to
the reader. 
\endproof

Therefore, with probability greater than $1-p_N$, $\sup(\Phi_{N,\b}^h - 
\E \Phi_{N,\b}^h(z)$
does not exceed one half of the lower bound of the deterministic part, 
which implies that
$$
\eqalign{
\int_{\|z-\tilde{r}^h\|\geq \d_N}e^{-\b N \Phi_{N,\b}^h(z)}\, dz 
&\leq e^{-\b N \E \Phi_{N,\b}^h(\tilde{r}^h)} \int_{\|z-\tilde{r}^h\|\geq \d_N}
        e^{- \b N \frac{C}{2} \|z - \tilde{r}^h\|_2^2}\, dz \cr
&\leq e^{-\b N \E \Phi_{N,\b}^h(\tilde{r}^h)} e^{- \b N \frac{C}{4}\d_N^2 K}.
}
\Eq(C.206)
$$

We now turn to the denominator in \eqv(C.202). The probability
 that the fluctuation term
exceeds an $\varepsilon > 0$ is bounded by Lemma \ver.5:
$$
q_N \equiv \P \Big[\sup_{\|z-\tilde{r}^h\| < \d_N} |f_N(z) - \E f_N(z)| 
        \geq \varepsilon \Big] \leq K_1 e^{-K_2 N} + C_1 \varepsilon^{-2}
        e^{-C_2\varepsilon^2 N}.
\Eq(C.207)
$$
Using the Taylor expansion of $\E \Phi_{N,\b}^h(z)$ about $\tilde{r}^h$ up to order 2,
with an error term of order 3, we get that with probability higher than $1-q_N$,
$$
\eqalign{
\int_{\|z-\tilde{r}^h\| < \d_N} e^{-\b N \Phi_{N,\b}^h(z)}\, dz 
&\geq e^{-\b N (\Phi_{N,\b}^h(\tilde{r}^h + C'' \d_N^3 + \varepsilon))}  
        \int_{\|z-\tilde{r}^h\| < \d_N}e^{-\b N C' \|z- \tilde{r}^h\|_2^2}\,dz \cr
&\geq e^{-\b N (\Phi_{N,\b}^h(\tilde{r}^h + C'' \d_N^3 + \varepsilon))} 
        KN^{-1/2}(1 - e^{-\b N \frac{C'}{2} \d_N^2}).
}
\Eq(C.208)
$$

Combining \eqv(C.206) and \eqv(C.208) gives
$$
 \frac{\int_{\|z -\tilde{r}^h\| \geq \d_N} e^{-\b N \Phi_{N,\b}^h(z)}\, dz}
        {\int_{\|z -\tilde{r}^h\| < \d_N} e^{-\b N \Phi_{N,\b}^h(z)}\, dz}
  \leq \widetilde{K} e^{-\b N(\frac{C}{2}\d_N^2 - \varepsilon - C'' \d_N^3)}
\Eq(C.209)
$$
with probability greater than $1 - (q_N+p_N)$. Choosing $\d_N = N^{-1/5}$, 
$\varepsilon = N^{-1/5}$, implies that $\sum_{N} (p_N + q_N)<\infty$. 
Applying the Borel-Cantelli Lemma then gives the statement of 
Proposition \ver.8.
\endproof

Theorem 1 is now obvious:

\proofof {Theorem 1}
Let $f$ be a bounded continuous function. Then
$$
\eqalign{
\QQ_{N,\b}^h (f) &= f(\tilde{r}^h \QQ_{N,\b}^h (\1_{\{\|z-\tilde{r}^h\| \leq
                        \d_N\}}) 
                + \QQ_{N,\b}^h ((f(\tilde{r}^h - f)\1_{\{\|z-\tilde{r}^h\| \leq
                        \d_N\}}) \cr
& \quad +   \QQ_{N,\b}^h (f \1_{\{\|z-\tilde{r}^h\| > \d_N\}}) .
}
\Eq(C.402)
$$
Taking the limit $N \uparrow \infty$, we can replace $\QQ_{N,\b}^h$ by 
  $\wt{\QQ}_{N,\b}^h$
and use Proposition \ver.8. Since $f$ is  bounded, the 
third term on the right-hand side of \eqv(C.402) converges to zero, and since 
it is continuous, the second term also  vanishes too. These statements
are true $\P$-a.s. 
Finally we let $b = \|h\|_2 \rightarrow 0$. Again by continuity of $f$,
$f(\tilde r^h)\rightarrow f(r^*(\cos\vartheta,\sin\vartheta))$.
This proves the Theorem.\endproof\endproof

\newpage
\chap{3. Uniqueness of extrema of certain gaussian processes.}3
\def\varth{\vartheta}

In the previous chapter we have seen that the measures $\wt\QQ_{N,\b}$
concentrate on a circle of radius $r^*$ at the places where the 
random function $g_N(\varth)$ takes its minimum. In this section we
will show that these sets degenerate to a single point, a.s. 
in the limit $N\uparrow\infty$. To do so we first prove a uniqueness 
theorem for the absolute minimum of a certain class of strongly correlated
Gaussian processes. Then we show convergence 
in distribution 
of $g_N(\varth)$ to such a process
and finally we show that this implies also the desired convergence in 
distribution of our measures. We begin with the following general result.

\proposition {\ver.1}
 {\it  Suppose $\chi(t)$ is a real stationary Gaussian process which is 
periodic with period $T$. Suppose furthermore that its 
covariance function $r(s,t) = r(s-t)$ is even, $\in C^{\infty}[0,T]$, 
and $r(\tau)$ is less than $r(0)$ for all $\tau \in (0,T)$. 
Then there exists an equivalent process $\eta(t)$ having almost 
surely infinitely differentiable sample paths. 
Moreover, the probability that there exist two or more maxima
with equal height in $[0,T)$ is zero.
}

\proof Without restricting the generality, we can assume 
that ${\E}\,[\chi(t)] = 0$ and $\kappa = 
{\E}\,[\chi(t)^2] = 1$. 

By its continuity properties, $r(\tau)$ can be expanded about the
origin as
$$
r(\tau) = 1 - \frac{\lambda_2}{2!}\tau^2 + O(\tau^4).
\Eq(U.1)
$$
The first assertion then follows from the following result
due to Cram\'{e}r and Leadbetter (see [CL]), 
chapter 9.2).

\lemma{\ver.2}{\it 
  Suppose that for some $a > 3$,
  $$
    r(\tau) = 1 - \frac{\lambda_2}{2} \tau^2 + O\left(
      \frac{\tau^2}{|\ln|\tau||^a} \right),
 \Eq(U.2)
 $$
where $\lambda_2$ is a constant. Then there exists a process
$\eta(t)$ equivalent to $\chi(t)$ and possessing, with probability
one, a continuous derivative $\eta'(t)$.
}

\proof See Cram\'{e}r/Leadbetter [CL].

It is easily checked that by \eqv(U.1), $r(\tau)$
satisfies the condition \eqv(U.2) in Theorem \ver.2, 
which proves the statements about continuity and existence
of a continuous derivative.

Consider now the process $\chi'(t)$. Its covariance function
$\tilde{r}(\tau)$ is given by $\tilde{r}(\tau) = - r''(\tau)$
(see for example Leadbetter et al. [LLR],
p.\ 161, chapter 7.6). Then it can be expanded about
the origin as
$$
   \tilde{r}(\tau) = \lambda_2 - \frac{\lambda_4}{2} \tau^2
    + O(\tau^4).
\Eq(U.8)
$$
Then $\tilde{r}(\tau)$ also verifies condition \eqv(U.2) in
Theorem \ver.2. Repeating this argument implies, together
with the Borel-Cantelli Lemma, that there exists an equivalent
process $\eta(t)$ having, with probability one, infinitely 
differentiable sample paths.

From now on, we assume that $\chi(t)$ itself has the above 
continuity properties. 
We want to find the probability that there are not
two maxima with equal height in $[0,T)$, i.e.\ 
$$
  {\P} \left[ \exists s,t \in T \times T:
   |s - t| \neq k T, |\chi(t) - \chi(s)| = 0, |\chi'(t)| =
   |\chi'(s)| = 0 \right] = 0.
\Eq(U.9)
$$
We first show that for any $\vartheta > 0$, 
$$ 
{\P} \left[ \exists s,t \in T \times T:
    \Big|k T - |s -t|\Big|\geq \vartheta, 
    |\chi(t) - \chi(s)| = 0, |\chi'(t)| =
     |\chi'(s)| = 0 \right] = 0
\Eq(U.10)
$$
Let us choose a collection of grid points $t_i \in T$, separated by 
some distance $\varepsilon > 0$. By the continuity properties, $\chi$ and 
$\chi'$ are Lipschitz-continuous with a.s.-finite constants $C_0$, $C_1$.
Consider the set $\tilde{\Omega}_C \sb \Omega$ such that 
$C_0$ and $C_1$ are bounded by some number $C>0$. Then, by 
Lipschitz-continuity, $\chi'(t) = 0$, $t \in [t_i, t_{i+1})$ implies that 
(for some $x \in [t_i,t]$)
$$
|\chi'(t_i)| \leq C \varepsilon.
\Eq(U.11)
$$
Similarly, $|\chi(t) - \chi(s)| = 0$ implies
$$
|\chi(t_i) - \chi(t_j)| \leq 2 C \varepsilon 
\Eq(U.11bis)
$$
where $t - t_i < \varepsilon$, $s - t_j < \varepsilon$. Then we can 
estimate the probability of the event in \eqv(U.10) (on $\tilde{\Omega}$)
by
$$
\eqalign{
&{\P} \left[ \exists s,t \in T \times T:
    \Big|k T - |s -t|\Big|\geq \vartheta, 
    |\chi(t) - \chi(s)| = 0, |\chi'(t)| =
     |\chi'(s)| = 0 \right] \cr
&\leq {\P} \Bigl[ \exists t_i,t_j: \big|k T - |s -t|\big|\geq \vartheta,
        |\chi(t_i)-\chi(t_j)| \leq 2 C \varepsilon, 
        |\chi'(t_i)| \leq C \varepsilon, \cr
&\quad |\chi'(t_j)| \leq C \varepsilon \Bigr].
}
\Eq(U.11ter)
$$
Let us denote the event  appearing on the left-hand side of \eqv(U.11ter)
by $\AA_{\vartheta}$, and the event appearing on the right-hand side by
$\BB_{\vartheta,\varepsilon}$.
The probability $\P[\BB_{\vartheta,\varepsilon}]$ can be estimated by the standard bound
$$
    {\P} \left[\BB_{\vartheta,\varepsilon}\right] 
    \leq\!\! \sum_{\left| kT - |t_i-t_j|\right|\geq \vartheta}
       {\P} \left[ 
       |\chi(t_i)-\chi(t_j)| \leq 2 C \varepsilon,
       |\chi'(t_i)| \leq C \varepsilon,
       |\chi'(t_j)| \leq C \varepsilon\right].
\Eq(U.12)
$$
Now, for any fixed $i,j$, 
$$
  \left(\chi(t_i) - \chi(t_j), \chi'(t_i),\chi'(t_j) \right)
\Eq(U.13)
$$
is a Gaussian vector, and due to the condition on $|t_i - t_j|$ and 
the assumption concerning $r(\tau)$, its distribution is non-degenerate. 
Therefore, each term in the sum on the right-hand side of \eqv(U.12) 
can be bounded by
$$
  {\P} \left[ 
       |\chi(t_i)-\chi(t_j)| \leq 2 C \varepsilon,
       |\chi'(t_i)| \leq C \varepsilon,
       |\chi'(t_j)| \leq C \varepsilon \right]
  \leq K \varepsilon^3 C^3 (2 \pi \sigma_{i,j})^{-1},
\Eq(U.14)
$$
where $\sigma_{i,j}$ is the determinant of the non-degenerate covariance 
matrix of the random vector \eqv(U.13). Since the $t_i, t_j$ are 
chosen in a compact set, this quantity can be bounded uniformly in $i,j$. 
We thus get
$$
 {\P} \left[ 
       |\chi(t_i)-\chi(t_j)| \leq 2 C \varepsilon,
       |\chi'(t_i)| \leq C \varepsilon,
       |\chi'(t_j)| \leq C \varepsilon\right]
  \leq K(\vartheta) \varepsilon^3 C^3.
\Eq(U.15)
$$
Finally, the number of allowed pairs $(i,j)$ in the sum in equation
\eqv(U.12) does not exceed  $T^2 \varepsilon^{-2}$, which implies 
that
$$\eqalign{
    {\P} \left[\AA_{\vartheta}\right]
    \leq& {\P}\left[\BB_{\vartheta,\varepsilon} \right]
      + {\P} \left[ \tilde{\Omega}^c_C \right] \cr
    \leq& K(\vartheta)T^2 \varepsilon^{-2} \varepsilon^3 + 
      {\P}\left[ \tilde{\Omega}^c_C \right],
}\Eq(U.16)
$$
keeping track of the set $\tilde{\Omega}^c_C$ on which the above estimates are 
not valid. Now choose $C=C(\varepsilon) = o(\varepsilon^{-1/3})$, and observe
that due to the continuity properties
$$
\eqalign{
\lim_{\varepsilon \rightarrow 0} {\P} \left[
        \tilde{\Omega}_{C(\varepsilon)}^c \right]
&= \P \left[ \bigcap_{n \in \N} \{C \geq n\} \right] \cr
&= 0.
}
\Eq(U.16bis)
$$
Finally, letting $\varepsilon$ tend to zero
in \eqv(U.16) gives that the probability \eqv(U.11) is zero. 

This shows that local maxima are separated  
with probablity one. In particular,
constant pieces and no accumulation points of maxima. 
This concludes its proof.
\endproof

\corollary{\ver.3}{\it
  Suppose $\chi(t)$ satisfies the conditions in Proposition
\ver.1. Then $\chi(t)$ has a.s.\ only one global
maximum in any interval $[s,s+t]$, $t<T$.
}

To see that Proposition \ver.1 is relevant for our problem, we will next 
show that the process $g_N(\varth)$ converges to a 
process of the type covered by this proposition. In fact we have

\proposition{\ver.4} {\it
  Let $g:{\R} \rightarrow {\R}^+$, $g \in 
C^{\infty}$ be an aperiodic even function. Suppose also that 
$\chi_i(\vartheta)$, $\vartheta \in [0, 2\pi]$ is the stochastic 
process given by
$$
    \chi_i(\vartheta) = g\left(r \zeta_i \cos(\vartheta - 
      \phi_i) \right),
\Eq(U.18)
$$
where $r$ is a positive constant, $\{\zeta_i\}_{i \in \N}$,
$\{\phi_i\}_{i \in \N}$ are two mutually independent families
of i.i.d.\ random variables, distributed as $cxe^{-x^2}$($\zeta_i$),
and uniformly ($\phi_i$). Then the process $\eta_N$ given by
$$
  \eta_N(\vartheta) \equiv \frac{1}{\sqrt{N}} \sum_{i=1}^N 
      \{\chi_i(\vartheta) - {\E}\,\chi_i(\vartheta)\}
\Eq(U.19)
$$
converges in distribution to a strictly stationary Gaussian process 
$\eta(\vartheta)$ having a.s.\ continuously differentiable sample
paths. Furthermore, $\eta(\vartheta)$ has a.s.\ only one 
global maximum on any interval $[s,s+t]$, $t<\pi$
}

\remark We will use this proposition of course with $g(\cdot)=
\ln\cosh(\b \cdot)$. Then the proposition implies that the 
process $g_N(\varth)-\E\, g_N(\varth)$ converges to a Gaussian process with 
the above properties.

\proof As $\xi_i(\varth)$ are i.i.d. stationary processes on the circle
which are infinitely differentiable, the convergence of the process to a 
stationary Gaussian process on the circle  is a simple 
application of the central limit theorem in Banach spaces (see e.g. [LT]). 
A computation shows that the covariance of the limiting  process is given by
$$
\eqalign{  
    f(s,t) 
      &= {\E}\left[
       \left(\chi_1(s) - \E\,\chi_1 (s) \right)
       \left(\chi_1(t) - \E\,\chi_1 (t) \right) \right] \cr
      &= {\E}\left[g\left(r\zeta_1\cos(\varphi_1)\right)
    g\left(r\zeta_1\cos(t-s -\varphi_1)\right)\right]-
\left({\E} \left[g\left(r\zeta_1\cos(\varphi_1)\right)
           \right]\right)^2 
}
\Eq(U.25)
$$
We see that this function is even, and
is in $C^{\infty}$ as a function of $\tau = t-s$. Moreover,
it is easily checked that the covariance function 
$f(\tau)$ is strictly smaller than $f(0)$, whenever $\tau \neq k \pi$.
Proposition \ver.1 and Corollary \ver.3
then imply the assertions about continuity and non-existence 
of more than one global maximum.
This concludes the proof of Proposition 
\ver.4. \endproof

We now check some intuitive properties of the position of 
the minimum of the Gaussian process from Proposition 
\ver.1 (for those $\omega$ such that the minimum exists
and is unique).

\proposition{\ver.5}  {\it 
  Suppose that the conditions of Proposition \ver.1 are satisfied.
  Define $(\Omega',{\cal F}', {\P}')$ to be the restriction
  of $(\Omega,{\cal F}, {\P })$ to all $\omega$ such that
  the conclusions of Proposition \ver.1 are true. Then the 
  position of the minimum
  $$
    \vartheta^*[\omega] \equiv 
\arg\min_{\varth\in[0,\pi)}\chi[\omega](\varth)
  \Eq(U.26)
$$
  of the sample path $\chi[\omega]$ is a random variable
  with uniform distribution on $[0,\pi)$.
}

\proof To prove that $\vartheta^*[\omega]$ is a random
variable, it is enough to show that for all intervals ${\cal U}
= (a,b) \subseteq [0,\pi)$, the set $\vartheta^*{}^{-1}({\cal U})$ is
in $\cal F'$. We note that by the continuity of $\chi$ on $[0,\pi)$ for
all $\omega \in \Omega'$, 
$$\eqalign{
    \vartheta^*{}^{-1}({\cal U}) &\equiv \{\omega \in \Omega:
      \chi[\omega](\cdot) \text{\rm  assumes its minimum in }
      {\cal U}\} \cr
    &= \{\omega \in \Omega': \exists t \in {\cal U} \cap 
      {\Q} \text{\rm  such that} \forall s \in
      {\cal U}^c \cap {\Q}, \chi(t) < \chi(s)\}.
}
\Eq(U.27)
$$
The second line can be written as
$$
    \bigcup_{t \in {\cal U} \cap {\Q}} 
    \bigcap_{s \in {\cal U}^c \cap {\Q}}
  \{\omega \in \Omega': \chi(t) < \chi(s)\},
\Eq(U.28)
$$
which clearly is in ${\cal F}'$.

Equation \eqv(U.28), together with the strict stationarity (since
it is a real stationary process) of the 
process $\chi$, implies the uniformity of the distribution.
This proves Proposition \ver.5. \endproof

Finally, to get some information about the convergence of functions
of the position of the minimum, we use the following two results.

\lemma{\ver.6}{\it
  Let $\PP([0,\pi))$ be the space of $T$-periodic, continuous functions, 
  having only one global minimum,
  together with the supremum norm. Then the position $\vartheta^*$ of the global
  minimum is a continuous function from ${\cal P}([0,\pi))$ to 
  $[0,\pi)$.
}

\proof Suppose that there exists a sequence of functions
$\{f_n\}$ converging to $f \in {\cal P}([0,\pi))$, such that the 
sequence of the global minima $\vartheta^*_n$ does not converge
to $\vartheta^*$, the global minimum of $f$. Then there exists
an $\varepsilon > 0$ and a subsequence $\{f_{n_k}\}$, such that
for all $n_k$, $|\vartheta^*_{n_k} - \vartheta^*| > \varepsilon$.

Now, since $\vartheta^*$ is the unique global minimum of $f$, $\exists
\d_{\varepsilon} > 0$ such that 
$$
f(\vartheta^*_{n_k}) > f(\vartheta^*) + \d_{\varepsilon}.
\Eq(U.101)
$$
Similarly, since $\vartheta^*_{n_k}$ is the unique minimum of $f_{n_k}$,
$\exists \d'_{\varepsilon,n_k} > 0$ such that
$$
f_{n_k}(\vartheta^*) > f_{n_k}(\vartheta^*_{n_k}) + \d'_{\varepsilon,n_k}.
\Eq(U.102)
$$
Furthermore, since $f_{n_k}$ converges in the supremum norm, $\forall \d > 0$, 
$\exists K_{\d} \in \N$ such that
$$
\forall \vartheta \in [0,\pi), \forall k>K_{\d}, \quad |f_{n_k}(\vartheta) - f(\vartheta)| < \d.
\Eq(U.103)
$$
For any $k>K_{\d}$ one can therefore write
$$
\eqalign{
f_{n_k}(\vartheta^*) - f(\vartheta^*) &= f_{n_k}(\vartheta^*) - f_{n_k}(\vartheta^*_{n_k})
        +  f_{n_k}(\vartheta^*_{n_k}) - f(\vartheta^*_{n_k}) +  f(\vartheta^*_{n_k})
        - f(\vartheta^*) 
\cr
&> \d'_{\varepsilon,n_k} - \d + \d_{\varepsilon} 
\cr
&> \d_{\varepsilon} - \d.
}
\Eq(U.104)
$$
Now choose $\d = \frac{1}{3}\d_{\varepsilon}$. Then for all $k > K_{\frac{1}{3}\d}$,
$$
f_{n_k}(\vartheta^*) - f(\vartheta^*) > \frac{2}{3} \d_{\varepsilon} > \d,
\Eq(U.105)
$$
which contradicts the assumption of uniform convergence. 
\endproof

%
%
%

The following result is crucial to link the weak convergence of 
the process $g_N(\varth)$ to the weak convergence of the 
measures $\QQ_{N,\b}$. 

\proposition {\ver.7} {\it Define the random sets
$$
L_N[\o] = \big\{ \varth \in [0,\pi): \eta_N[\o](\varth) - \min_{\varth'} \eta_N[\o](\varth') \leq \varepsilon_N \big\}
\Eq(U.41)
$$
with $\varepsilon_N$ some sequence converging to zero. Then 
$$
L_N\limlaw \varth^*
\Eq(U.42)
$$
}

\proof Using the 
{\it method of a single probability space} (see [Shi], Chapter 3,
Section 8, Theorem 1) one can construct  a probability space $(\Omega^*, \FF^*,
\P^*)$ and random processes $\eta^*_N$, $\eta^*$, such that
$$
\eta^*_N \rightarrow \eta^*, \quad \P^*-a.s.,
\Eq(U.39)
$$
and
$$
\eta^* \,\,{\buildrel \DD \over =}\,\, \eta, \qquad \eta^*_N \,\,{\buildrel \DD \over =}
               \,\,\eta_N.
\Eq(U.40)
$$
Now introduce the random level sets
$$
\eqalign{
L^*_N[\o^*] &= \big\{ \varth \in [0,\pi): \eta^*_N[\o^*](\varth) - 
\min_{\varth'}\eta^*_N[\o^*](\varth') \leq
        \varepsilon_N \big\},
}
$$
Then $L_N$ and  
$L^*_N$ have the same 
distribution. But since $\eta_N^*[\o]$ converges almost surely
to $\eta^*[\o]\in \PP([0,\pi))$, one sees that due to Lemma \ver.6
 $L^*_N[\o]$ converges $\P^*$-a.s. to the position of the unique absolute 
minimum of
$\eta^*[\o^*]$. But this minimum has the same distribution 
as that of $\eta$, which is the uniform distribution by Proposition \ver.5.
 Therefore, $L_N$ converges in distribution to a 
uniformly distributed point
on $[0,\pi)$. \endproof 

We have finally all tools available to prove Theorem 3.

\proofof {Theorem 3} We have to check convergence on the following type
of functions $F:\MM(\R^2) \rightarrow \R$
$$
F(\mu) = \wt{F}(\mu(f_1),\ldots,\mu(f_k)),
\Eq(C.301)
$$
where $\wt{F}$ is a polynomial function, and $f_1,\ldots,f_k$ are bounded
continuous functions from $\R^2 \rightarrow \R$. Convergence in law then means
that
$$
\lim_{N \uparrow \infty} \E \bigg[ F(\QQ_{N,\b}[\o]) \bigg] 
        = \frac{1}{\pi} \int_{0}^{\pi} 
        F(\frac{1}{2}\d_{(m^* \cos \vartheta, m^* \sin \vartheta)} +
        \frac{1}{2}\d_{(m^* \cos \vartheta+\pi, m^* \sin \vartheta+\pi)})
        \, d\vartheta.
\Eq(C.302)
$$
The left-hand side of \eqv(C.302) is explicitly written as
$$
\lim_{N \uparrow \infty} \E \bigg[ \wt{F}(\QQ_{N,\b}[\o](f_1),\ldots,
        \QQ_{N,\b}[\o](f_k))\bigg].
\Eq(C.303)
$$
We now treat the individual arguments of $\wt{F}$ in \eqv(C.303). Let $A_N[\o]$
(the level sets in the previous lemmata) 
be decomposed into its $2 l'$ connected
components $A_{N,j_N}[\o]$. As a consequence of  
Lemma \ver.7, there exists $N[\o]$ which is finite a.s. such that  for all  
$N \geq N(\o)$, $l=1$,  and the two corresponding connected
components are symmetric with respect to the origin. 
Now choose arbitrary points
$x_{N,j_N}[\o] \in A_{N,j_N}[\o]$. Then we can decompose
$$
\eqalign{
\wt{\QQ}_{N,\b}[\o](f_i) &= \sum_{j_N} f_i(x_{N,j_N}) \wt{\QQ}_{N,\b}[\o] (\1_{A_{N,j_N}})
        + \sum_{j_N} \wt{\QQ}_{N,\b}(\1_{A_{N,j_N}} 
        (f_i(x_{N,j_N}) - f_i))  \cr
&\quad + \wt{\QQ}_{N,\b}(\1_{A_N^c} f_i).
}
\Eq(C.304)
$$
Expanding $\wt{F}$ using the decomposition \eqv(C.304),  we get a sum consisting
of two different types of terms: (i), summands that are products of the first sum on the right-hand
side of \eqv(C.304) only, and (ii), summands where at least one of the second and third term
from the right-hand side of \eqv(C.304) enter. Proposition 2.3 and 
 Proposition 3.7, and the continuity and boundedness of the $f_i$'s imply that the terms
of type (ii) vanish $\P$-a.s., as $N \uparrow \infty$. In the limit, the only terms left are of type 
(i), which together sum up to
$$
\wt{F}\left(\sum_{j_N} f_1(x_{N,j_N})\wt{\QQ}_{N,\b}[\o] (\1_{A_{N,j_N}}),
        \ldots, \sum_{j_N} f_k(x_{N,j_N})\wt{\QQ}_{N,\b}[\o] (\1_{A_{N,j_N}}))
\right.
\Eq(C.305)
$$
All arguments of $\wt{F}$ in \eqv(C.305) converge in distribution to
$$
\frac{1}{2}f_i((m^* \cos\vartheta,m^* \sin \vartheta)) + 
        \frac{1}{2} f_i((m^* \cos\vartheta + \pi,m^* \sin \vartheta + \pi)),\quad \forall i =1,\ldots,k
\Eq(C.306)
$$
where $\vartheta$ is a uniformly distributed r.v.\ on $[0,\pi)$,
by Proposition 3.7. But convergence in distribution means
by definition that
$$
\eqalign{
\lim_{N \uparrow \infty} & \E \bigg[
        \wt{F}(\sum_{j_N} f_1(x_{N,j_N})\wt\QQ_{N,\b}[\o] (A_{N,j_N}),
        \ldots, \sum_{j_N} f_2(x_{N,j_N})\wt\QQ_{N,\b}[\o] (A_{N,j_N})) \bigg]
\cr
&= \frac{1}{\pi} \int_{0}^{\pi}  
        \wt{F}(\frac{1}{2}f_i((m^* \cos\vartheta,m^* \sin \vartheta)) + 
        \frac{1}{2} f_i((m^* \cos\vartheta + \pi,m^* \sin \vartheta + \pi))\,d\vartheta,
}
\Eq(C.307)
$$
which in turn is by definition equal to
$$
\frac{1}{\pi}\int_{0}^{\pi}
        F(\frac{1}{2}\d_{(m^* \cos\vartheta,m^* \sin \vartheta)} +
        \frac{1}{2}\d_{(m^* \cos\vartheta + \pi,m^* \sin \vartheta + \pi)})\, d\vartheta.
\Eq(C.308)
$$
This proves the convergence  in law \eqv(I.12) in Theorem 3. To obtain
the identification  of the metastate, just note that the process 
$\eta_N(\vartheta)[\o]$ actually converges to the same Gaussian process
under any of the conditional laws $\P[\cdot|\FF_n]$, where $\FF_n$ is the 
sigma-algebra generated by the random variables $\xi_i,i\leq n$.   
\endproof\endproof

\vskip2cm

\chap{4. Volume dependence,  empirical metastates, superstates}4

We conclude this paper with the discussion of some more sophisticated concepts 
that have been proposed by Newman and Stein [NS2] and Bovier and Gayrard
[BG3] and that should capture in more detail the actual asymptotic 
volume dependence of the Gibbs measures. 
In fact, the first question one may ask is whether for a fixed realization
 as the volume grows the
finite volume Gibbs states really explore all the possibilities in the 
support of the metastate. One way of stating that this is the case is the 
following

\theo{\ver.1} {\it There exist (deterministic) sequences $N_k\uparrow\infty$
such that the empirical metastate 
$$
\frac 1k\sum_{\ell=1}^k \d_{\QQ_{N_k,\b}}, 
\Eq(4.100)
$$
converges almost surely to the law of $\QQ_{\infty,\b}$.
}

\proof We have seen that the measure $\QQ_{N_k,\b}$ is sharply concentrated
on the circle of radius $r^*$ and at the angle where the process 
$g_{N_k}(\varth)$ (defined in \eqv(C.101) takes its absolute minimum. 
The idea is to choose $N_k$ in such a way that these angles will be 
virtually independent for different $k$.
Now note that we can write
$$
g_{N_k}(\varth) =\wt g_{k}(\varth) + R_{k}(\varth),
\Eq(4.101)
$$
where
$$
\wt g_{k}(\varth)=\frac 1{N_k} \sum_{i=N_{k-1}+1}^{N_k}\ln\cosh(\b(r^*\z_i
\cos(\varth-\varphi_i))),
\Eq(4.102)
$$ 
are independent for different $k$ by construction and
$$
 R_{k}(\varth)=\frac1{N_k} \sum_{i=1}^{N_{k-1}}\ln\cosh(\b(r^*\z_i
\cos(\varth-\varphi_i))).
\Eq(4.103)
$$
Now by standard estimates identical to those presented in Section 3, one 
shows easily that there is a constant $C<\infty$ such that 
$$
\P\left[sup_{\varth\in[0,\pi)} |R_{k}(\varth)-\E R_k(\varth)|
\geq x \frac {N_{k-1}}{N_k} \right]
\leq C\exp\left(-x^2/C\right). 
\Eq(4.104)
$$
Thus we can always choose $N_k$ growing sufficiently rapidly (e.g. $N_k=k!$)
 such that 
$R_k$ is totally negligible compared to $\wt g_k$ for large $k$, and the 
position of the absolute minimum of $g_{N_k}(\varth)$ is asymptotically equal
to that of $\wt g_k(\varth)$. This allows us to 
approximate for large $k$ the random measures $\d_{\QQ_{N_k,\b}}$
by independent measures and from this the asserted result follows from the 
law of large numbers. \endproof

\remark Theorem \ver.1 says that that the empirical metastate constructed with
sparse subsequences converges to the Aizenman-Wehr metastate, a.s.. This
is a special example of a general theorem due to Newman and Stein [NS2]
(where however they require possibly subsequences $\ell_i$ in the definition
\eqv(4.100)).

Rather than considering the empirical metastate with sparse subsequences
one may be interested in the volume dependence as the volume grows at its 
natural pace. To capture this, the idea put forward in [BG3] is to 
construct a measure valued stochastic process
$$
\mu^t_\b\equiv \lim_{N\uparrow\infty} \mu_{\b,[tN]},
\Eq(4.1)
$$
with $t\in (0,1]$ and to consider either the (conditional)
probability distribution of this process (the ``superstate'' [BG3]) or the
(conditional) empirical distribution of the process (the ``empirical 
metastate'' [NS2]). Let us see what this entails in our 
context. The reader who has been following the 
exposition of the last two chapters will easily be convinced that this problem 
amounts to study the quantity 
$$
\varth(t)\equiv \arg\min_{\th\in [0,\pi)}\left(\chi_t(\th)\right),
\Eq(4.2)
$$
where $\chi_t(\th)$ is the distributional limit
of the process
$$
\chi^t_N(\varth)\equiv g_{[tN]}(\varth)-\E g_{[tN]}(\varth).
\Eq(4.3)
$$
where $g_N(\th)$ is defined in \eqv(C.101). By completely standard arguments 
one shows that the following invariance principle holds:

\lemma {\ver.2} {\it The process $\chi^t_N(\varth)$
converges in distribution, as $N\uparrow\infty$ to the 
Gaussian process $\chi_t(\varth)$, $t\in (0,1],\varth\in [0,\pi)$ with 
mean zero and covariance 
$$
C(\varth,\varth',t,t')\equiv \frac {t\wedge t'}{\sqrt{tt'}}f(\varth,\varth'),
\Eq(4.4)
$$
where 
$$
f(\varth,\varth')=
 {\E}\left[\ln\cosh\left(\b r\zeta_1\cos(\varphi)\right)
         \ln\cosh\left(\b r\zeta_1\cos\left(\varphi-(\varth-\varth')\right)
\right)\right].
\Eq(4.5)
$$
}

$\chi_t(\th)$ is a rather curious Gaussian process: as a function
of $t$, to fixed $\varth$ it is (normalized) Brownian motion, while for
fixed $t$ as a function of $\varth$ it is the $C^\infty$ process discussed in 
the previous section. The question is then what can be said about the 
process $\varth_t$, defined by \eqv(4.2)?  

Some facts follow easily. For instance, the process is almost surely 
single valued
for all $t\in (0,1]$ except possibly on some Cantor set of zero 
Lebesgue measure. 
On the other hand, it seems natural that such an exceptional set will exist and
that a typical realization will have continuous pieces and ``jumps''. 
Also, for $t$ going to zero, the process 
``circles'' around rapidly since $ \chi_t$ and $\chi_s$ become uncorrelated as
$s\downarrow 0$. But otherwise we do not see any immediate more specific 
characterization of the process or its path-properties. 

\vskip2cm
{\headline={\ifodd\pageno\rightheadline \else \leftheadline \fi}}
\def\rightheadline{\it  {References}\hfil\tenrm\folio}
\def\leftheadline{\tenrm \folio \hfil\it  {Gaussian Hopfield}}

\chap{References}1

\item{[AW]} M.~Aizenman, J.~Wehr, {\it Rounding effects of quenched randomness 
of first-order phase transitions}, Comm. Math. Phys. {\bf 130}, 489--528, 1990.
\item{[CL]} H.~Cram\'er, M.~R.~Leadbetter,
  {\it Stationary and Related Stochastic Processes},
  Wiley, 1967.
\item{[BGP]} A.~Bovier, V.~Gayrard, P.~Picco,
  {\it Gibbs States of the Hopfield Model in the Regime of Perfect Memory},
  PTRF {\bf 100}, 329--363, 1994.
\item{[BG1]} A.~Bovier, V.~Gayrard,
  {\it Hopfield Models as Generalized Random Mean Field Models}, in
  {\it Mathematical Aspects of Spin Glasses and Neural Networks},
  A.~Bovier and P.~Picco (eds.), Progress in Probability, Birkh\"auser,
  1998.
\item{[BG2]} A.~Bovier, V.~Gayrard,
  {\it The Retrieval Phase of the Hopfield Model: A Rigorous Analysis of
    the Overlap Distribution}, PTRF {\bf 107}, 61--98, 1997.
\item{[BG3]} A.~Bovier, V.~Gayrard, {\it Metastates in the Hopfield model 
in the replica symmetric regime}, MPAG  {\bf 1}, ??-??, 1998.
\item{[BF]} A.~Bovier, J.~Fr\"ohlich, {\it A heuristic theory of the 
spin-glass phase}, J. Stat. Phys. {\bf 44}, 347--391, 1986.
\item{[vE]} A.~C.~D. van Enter, {\it Stiffness exponent, number of pure states,
and Almeida-Thouless line in spin-glasses} J.\ Stat.\ Phys. {\bf 60}, 275-279,
1990.
\item{[FH1]} D.~S.~Fisher, D.~A.~Huse, {\it Ordered phase of 
short-range spinglasses}, Phys. Rev. Lett. {\bf 56}, 1601--1604, 1986.
\item{[FH2]} D.~S.~Fisher, D.~A.~Huse, {\it Equilibrium behavior of the
spin-glass ordered phase}, Phys. Rev. B {\bf 38}, 386--411, 1988.
\item{[FH3]} D.~S.~Fisher, D.~A.~Huse, {\it Pure states in spin glasses},
J. Phys. A {\bf 20}, L997-1003, 1987.
\item{[FH4]} D.~S.~Fisher, D.~A.~Huse, {\it Absence of many states in realistic spin glasses}, 
J. Phys. A {\bf 20}, L1005--1010, 1987.
\item{[Ge]} S.~Geman,
  {\it A Limit Theorem for the Norms of Random Matrices},
  Ann.\ Prob. {\bf 8}, 252--261, 1980.
\item{[Ku1]} C.~K\"ulske, {\it Metastates in disordered mean field models:
Random field and Hopfield models}, J. Stat. Phys. {\bf 88}, 1257--1293, 1997.
\item{[Ku2]} C.~K\"ulske, {\it Limiting Behavior of Random Gibbs Measures:
Metastates in Some Disordered Mean Field Models}, in {\it Mathematical 
Aspects 
of Spin Glasses and Neural Networks}, A.~Bovier, P.~Picco (eds.), 
Progress in Probability 41, Birkh\"auser, Boston-Basel-Berlin, 1998.
\item{[LLR]} M.~R.~Leadbetter, G.~Lindgren, H.~Rootz\'en,
  {\it Extremes and Related Properties of Random Sequences
    and Processes},
  Springer, Berlin-Heidelberg-New York, 1983.
 \item{[LT]} M. Ledoux and M. Talagrand, ``Probability in Banach spaces'',
 Springer, Berlin-Heidelberg-New York, 1991.
\item{[MPR]} E.~Marinari, G.~Parisi, J.~J.~Ruiz-Lorenzo,
{\it Numerical Simulations of Spin Glass Systems}, in {\it Spin Glasses and
Random Fields}, A.~P.~Young (ed.), World Scientific, 1998. 
\item{[MPV]} M.~M\'ezard, G.~Parisi, M.~A.~Virasoro,
{\it Spin Glass Theory and Beyond}, World Scientific, 1987.
\item{[N]} C.~M.~Newman, {\it Topics in Disordered Systems}, Birkh\"auser,
 Boston-Basel-Berlin, 1997.
\item{[NS1]} C.~M.~Newman, D.~L.~Stein, {\it Non-mean-field behavior of
realistic spin glasses} Phys. Rev. Lett. {\bf 76}, 515-518, 1996.
\item{[NS2]} C.~M.~Newman, D.~L.~Stein, {\it Thermodynamic Chaos and 
the Structure of Short-Range Spin Glasses}, in {\it Mathematical Aspects of
Spin Glasses and Neural Networks}, A.~Bovier and P.~Picco (eds.), Progress in
Probability 41, Birkh\"auser, Boston-Basel-Berlin, 1998.
\item{[NS3]} C.~M.~Newman, D.~L.~Stein, {\it Spatial inhomogeneity 
and thermodynamic chaos}, Phys. Rev. Lett. {\bf 76}, 4821--4824, 1996.
\item{[NS4]} C.~M.~Newman, D.~L.~Stein, {\it Metastate approach to 
thermodynamic chaos}, Phys. Rev. E {\bf 55}, 5194--5211, 1997.
\item{[NS5]} C.~M.~Newman, D.~L.~Stein, {\it Multiple states and 
thermodynamic limits in short ranged Ising spin glass models}, Phys. Rev. B
{\bf46}, 973--982, 1992.
\item{[NS6]} C.~M.~Newman, D.~L.~Stein, {\it Simplicity of state and 
overlap structure in finite-volume realistic spin glasses}, Phys. Rev. E 
{\bf 57}, 1356--1369, 1998. 
\item{[Shi]}A.~N.~Shiryaev,
  {\it Probability},
  2nd ed., GTM 95,
  Springer, Berlin-Heidelberg-New York, 1996.

\end